\documentclass[11pt]{article}

\include{setup}

\usepackage{graphics}
\usepackage{cite}
\usepackage{axodraw}
\usepackage{rotating}
\usepackage{url}
\usepackage{epsfig}

\makeatletter
\if@twoside \oddsidemargin -17pt \evensidemargin 00pt
\else \oddsidemargin 00pt \evensidemargin 00pt
\fi
\def\paragraph{\@startsection{paragraph}{4}{\z@}{+2.00ex plus
 +1ex minus +.2ex}{1.5ex plus .2ex}{\it\normalsize}}

\expandafter\ifx\csname mathrm\endcsname\relax\def\mathrm#1{{\rm #1}}\fi

\renewcommand{\theequation}{\thesection.\arabic{equation}}
\newcounter{saveeqn}

\@addtoreset{equation}{section}

\makeatother

\def\beq{\begin{equation}}
\def\eeq{\end{equation}}
\def\beqar{\begin{eqnarray}}
\def\eeqar{\end{eqnarray}}
\def\barr#1{\begin{array}{#1}}
\def\earr{\end{array}}
\def\bfi{\begin{figure}}
\def\efi{\end{figure}}
\def\btab{\begin{table}}
\def\etab{\end{table}}
\def\bce{\begin{center}}
\def\ece{\end{center}}
\def\nn{\nonumber}

\def\text{\textstyle}

\def\ga{\gamma}

\def\Ga{\Gamma}
\def\De{\Delta}

\def\refeq#1{\mbox{(\ref{#1})}}
\def\reffi#1{\mbox{Figure~\ref{#1}}}

\def\refta#1{\mbox{Table~\ref{#1}}}

\def\refse#1{\mbox{Section~\ref{#1}}}
\def\refses#1{\mbox{Sections~\ref{#1}}}

\def\citere#1{\mbox{Ref.~\cite{#1}}}
\def\citeres#1{\mbox{Refs.~\cite{#1}}}

\newcommand{\GeV}{\unskip\,\mathrm{GeV}}

\newcommand{\TeV}{\unskip\,\mathrm{TeV}}

\def\mathswitch#1{\relax\ifmmode#1\else$#1$\fi}
\def\mathswitchr#1{\relax\ifmmode{\mathrm{#1}}\else$\mathrm{#1}$\fi}
\def\mathswitchit#1{\relax\ifmmode{#1}\else$#1$\fi}

\newcommand{\PW}{\mathswitchr W}
\newcommand{\PZ}{\mathswitchr Z}

\newcommand{\Pg}{\mathswitchr g}
\newcommand{\PH}{\mathswitchr H}

\newcommand{\Pe}{\mathswitchr e}

\newcommand{\Pd}{\mathswitchr d}

\newcommand{\Pu}{\mathswitchr u}
\newcommand{\Ps}{\mathswitchr s}
\newcommand{\Pc}{\mathswitchr c}
\newcommand{\Pb}{\mathswitchr b}
\newcommand{\Pt}{\mathswitchr t}
\newcommand{\Pp}{\mathswitchr p}

\newcommand{\Pep}{\mathswitchr {e^+}}
\newcommand{\Pem}{\mathswitchr {e^-}}

\newcommand{\PWp}{\mathswitchr {W^+}}
\newcommand{\PWm}{\mathswitchr {W^-}}

\newcommand{\MW}{\mathswitch {M_\PW}}

\newcommand{\MZ}{\mathswitch {M_\PZ}}
\newcommand{\MH}{\mathswitch {M_\PH}}

\newcommand{\Mb}{\mathswitch {m_\Pb}}
\newcommand{\Mt}{\mathswitch {m_\Pt}}
\newcommand{\GW}{\mathswitch {\Gamma_\PW}}
\newcommand{\GZ}{\mathswitch {\Gamma_\PZ}}

\newcommand{\scrs}{\scriptscriptstyle}
\newcommand{\sw}{\mathswitch {s_{\scrs\PW}}}
\newcommand{\cw}{\mathswitch {c_{\scrs\PW}}}

\newcommand{\ord}{{\cal O}}

\newcommand{\ri}{{\mathrm{i}}}

\newcommand{\rT}{{\mathrm{T}}}

\newcommand{\rd}{{\mathrm{d}}}

\newcommand{\sumav}{{\overline{\sum}}}

\newcommand{\EW}{{\mathrm{EW}}}
\newcommand{\QCD}{{\mathrm{QCD}}}

\newcommand{\OS}{{\mathrm{OS}}}

\newcommand{\LO}{{\mathrm{LO}}}
\newcommand{\NLO}{{\mathrm{NLO}}}

\def\Re{\mathop{\mathrm{Re}}\nolimits}

\newcommand{\spp}{\mathswitch {s_{\Pp\Pp}}}

\def\draftdate{\relax}
\def\mda{\relax}
\def\mua{\relax}
\def\mla{\relax}
\def\draft{
\def\thtystars{******************************}
\def\sixtystars{\thtystars\thtystars}
\typeout{}
\typeout{\sixtystars**}
\typeout{* Draft mode!
         For final version remove \protect\draft\space in source file *}
\typeout{\sixtystars**}
\typeout{}
\def\draftdate{\today}
\def\mua{\marginpar[\boldmath\hfil$\uparrow$]%
                   {\boldmath$\uparrow$\hfil}%
                    \typeout{marginpar: $\uparrow$}\ignorespaces}
\def\mda{\marginpar[\boldmath\hfil$\downarrow$]%
                   {\boldmath$\downarrow$\hfil}%
                    \typeout{marginpar: $\downarrow$}\ignorespaces}
\def\mla{\marginpar[\boldmath\hfil$\rightarrow$]%
                   {\boldmath$\leftarrow $\hfil}%
                    \typeout{marginpar: $\leftrightarrow$}\ignorespaces}
\def\Mua{\marginpar[\boldmath\hfil$\Uparrow$]%
                   {\boldmath$\Uparrow$\hfil}%
                    \typeout{marginpar: $\Uparrow$}\ignorespaces}
\def\Mda{\marginpar[\boldmath\hfil$\Downarrow$]%
                   {\boldmath$\Downarrow$\hfil}%
                    \typeout{marginpar: $\Downarrow$}\ignorespaces}
\def\Mla{\marginpar[\boldmath\hfil$\Rightarrow$]%
                   {\boldmath$\Leftarrow $\hfil}%
                    \typeout{marginpar: $\Leftrightarrow$}\ignorespaces}
\overfullrule 5pt
\oddsidemargin -15mm
\marginparwidth 29mm
}

\makeatletter

\def\eqnarray{\stepcounter{equation}\let\@currentlabel=\theequation
\global\@eqnswtrue
\global\@eqcnt\z@\tabskip\@centering\let\\=\@eqncr
$$\halign to \displaywidth\bgroup\hskip\@centering
  $\displaystyle\tabskip\z@{##}$\@eqnsel&\global\@eqcnt\@ne
  \hskip 2\arraycolsep \hfil${##}$\hfil
  &\global\@eqcnt\tw@ \hskip 2\arraycolsep $\displaystyle\tabskip\z@{##}$\hfil
   \tabskip\@centering&\llap{##}\tabskip\z@\cr}
\def\appendix{\par
 \setcounter{section}{0} \setcounter{subsection}{0}
 \def\thesection{\Alph{section}}}

\newcommand{\lsim}
{\;\raisebox{-.3em}{$\stackrel{\displaystyle <}{\sim}$}\;}

\def\dsl{\mathpalette\make@slash}
\def\make@slash#1#2{\setbox\z@\hbox{$#1#2$}%
  \hbox to 0pt{\hss$#1/$\hss\kern-\wd0}\box0}

\newcommand{\mr}[1]{{\mathrm{#1}}}

\makeatother

\newcommand{\mvev}{\nu_\mu\mu^+ \Pe^-\bar\nu_\Pe}
\newcommand{\mc}{\mathcal}
\def\non{\nonumber}
\newcommand{\muf}{\mu_\mr{F}}
\newcommand{\hs}{\hspace*}

\newcommand{\ed}{\end{document}}
\newcommand{\ffp}{\mathswitch{\mathrm{f\/f}'}}
\newcommand{\mfp}{\mathswitch{\mathrm{mf}'}}
\newcommand{\mmp}{\mathswitch{\mathrm{mm}'}}

\begin{document}
\thispagestyle{empty}
\def\thefootnote{\fnsymbol{footnote}}
\setcounter{footnote}{1}
\null
\draftdate\hfill FR-PHENO-2013-009 \\
\strut\hfill  MITP/13-060
\vfill
\begin{center}
{\Large \bf\boldmath
Next-to-leading order electroweak corrections to 
\\[.5em]
\boldmath{$\Pp\Pp\to\PWp\PWm\to4\,$}leptons at the LHC 
\\[.5em]
in double-pole approximation
\par} \vskip 2em
\vspace{.5cm}
{\large
{\sc M.\ Billoni$^1$, S.\ Dittmaier$^2$, B.\ J\"ager$^{1}$ and
C.\ Speckner$^2$} } 
\\[.5cm]
$^1$ {\it PRISMA Cluster of Excellence \& Institute of Physics, Johannes Gutenberg University, \\
55099 Mainz, Germany}
\\[0.3cm]
$^2$ {\it Physikalisches Institut, Albert-Ludwigs-Universit\"at Freiburg,\\
79104 Freiburg, Germany}
\par 
\end{center}\par
\vskip 2.5cm {\bf Abstract:} \par
We present the calculation of next-to-leading-order electroweak corrections to
$\PW$-boson pair production at the LHC, taking off-shell effects of the $\PW$~bosons and their
leptonic decays into account in the framework of the so-called double-pole approximation.
In detail, the lowest-order cross section and the photonic bremsstrahlung are 
based on full matrix elements with four-fermion final states, but the virtual one-loop corrections
are approximated by the leading contributions of a systematic expansion about the 
resonance poles of the two $\PW$~bosons. This expansion classifies the virtual corrections
into factorizable and non-factorizable corrections, the calculation of which is described in detail.
Corrections induced by photons in the initial state, i.e.\ 
photon--photon and quark--photon collision channels, are included and based on complete
matrix elements as well.
Our numerical results, which are presented for realistic acceptance cuts applied to
the $\PW$-boson decay products, qualitatively confirm recent results obtained for on-shell
$\PW$~bosons and reveal electroweak corrections of the size of tens of percent
in the TeV range of transverse momenta and invariant masses. 
In general, photon--photon 
and quark--photon induced contributions amount to $5{-}10\%$ of the full differential result.
Compared to previous
predictions based on stable W~bosons electroweak corrections, however, can change by
several percent because of realistic cuts on the W-boson decay products and corrections to the decays.
\par
\vskip 1.5cm
\noindent
October 2013
\null
\setcounter{page}{0}
\clearpage
\def\thefootnote{\arabic{footnote}}
\setcounter{footnote}{0}

%
%
\section{Introduction}
\label{se:intro}
%
Weak-boson pair production processes at high-energy colliders provide unique means for probing the electroweak sector of the Standard Model (SM). As they are sensitive to triple weak-gauge-boson vertices, these processes allow for stringent tests of the non-abelian structure of the electroweak interaction. Any deviations of measured cross sections and kinematic distributions from the values predicted by the SM
may point to ``new physics'' that manifests itself by an anomalous form of the gauge interactions. 
To unambiguously identify such deviations and clearly distinguish them from unknown perturbative corrections, calculations at high precision are essential. 
This task is of particular interest in the light of recent data
collected by the ATLAS~\cite{ATLAS:2012mec}  and CMS~\cite{Chatrchyan:2013oev} experiments
at the CERN Large Hadron Collider (LHC),
revealing some tension between the measured production cross section for W-boson pairs and its current theoretical prediction within the Standard Model.

In the past decades, tremendous effort went into the calculation of QCD corrections beyond the leading order (LO) to weak-boson pair production processes at hadron colliders, $\Pp\Pp\to VV$, 
which is dominated by quark--antiquark-annihilation subprocesses, $q\bar q\to VV$. 
Here and in the following, $V$ denotes a $\PW^\pm$ or a $\PZ$ boson. 
The next-to-leading order (NLO) QCD calculations for on-shell gauge-boson pair production processes have been presented in \citere{Ohnemus:1990za}, 
followed by calculations including the leptonic decays of the weak bosons in \citere{Baur:1995uv}. 
The latter have been implemented in the Monte Carlo program {\tt MCFM}~\cite{Campbell:1999ah}. More recently, 
gauge-boson pair production processes have also been matched to parton-shower programs at NLO QCD, 
first without~\cite{Frixione:2002ik}, 
later including leptonic decays~\cite{Hamilton:2010mb}. At the next-to-next-to-leading order (NNLO) in QCD only partial results for the two-loop~\cite{Chachamis:2008yb} and the one-loop squared~\cite{Chachamis:2008xu} virtual corrections in the high-energy limit exist. 
Soft-gluon resummation effects have first been assessed in \citere{Grazzini:2005vw} 
and have very recently been revisited in the framework of soft-collinear effective theory in \citere{Dawson:2013lya}. 
In both approaches, threshold logarithms were found to modify inclusive observables only mildly, while  more pronounced effects may emerge when hard selection cuts are applied. 
Apart from $q\bar q$-initiated production, gauge-boson pairs can also stem from loop-mediated gluon-scattering processes, $\Pg\Pg\to VV$. Although formally of higher order in the strong coupling, due to the large gluon luminosity at high-energy hadron colliders, this class of reactions contributes a non-negligible number of events to the full $VV$ production rate. 
Gluon-induced gauge-boson pair production processes have first been considered in
\citere{Dicus:1987dj}, with refined calculations being presented in \citere{Binoth:2005ua}. 
The most comprehensive QCD-based prediction for (off-shell) weak-diboson production at the LHC has been recently presented in 
\citere{Cascioli:2013gfa} in a study of these reactions as background to Higgs-boson production.
There the $q\bar q$ channels were merged at NLO QCD to a parton shower and combined with NLO QCD matrix elements
for hard jet emission; likewise this was done for the gluon--gluon channels at LO accuracy.

While QCD effects certainly represent the dominant source of perturbative corrections, a significant impact is expected also from electroweak (EW) corrections in the energy domain accessible at the 
LHC. EW corrections typically increase logarithmically with energy, and may reach several tens of percent at the 
LHC in the TeV range~\cite{Fadin:1999bq}.
Because of their strong energy dependence they affect the normalization of cross sections and, most importantly, 
also kinematic distributions. Only a precise knowledge of the EW corrections ensures that distortions in shape due 
to higher-order perturbative corrections can unambiguously be distinguished from the effects of physics beyond the SM. 
In \citeres{Accomando:2004de,Kuhn:2011mh} EW corrections to W-pair production at hadron colliders have been
estimated at NLO and NNLO by logarithmic approximations, respectively, confirming the global statements about their size
made above.
However, approximations based solely on EW high-energy logarithms
do not provide sufficient accuracy
in general, in particular, since those approximations are typically derived in the so-called Sudakov regime
(where both Mandelstam variables $\hat{s}$ and $|\hat{t}|$ are much larger than $\MW^2$), which is not necessarily the dominant 
kinematical domain in the high-energy limit. For instance, in the case of W-pair production
the bulk of the cross section
originates from the Regge limit (where $\hat{s}$ is large, but not $|\hat{t}|$). 
An assessment of EW corrections that goes beyond a qualitative level, thus, calls at least 
for complete NLO calculations beyond logarithmic approximations.

NLO EW corrections have recently been calculated for the on-shell production of weak-gauge boson pairs at 
hadron colliders~\cite{Bierweiler:2012kw,Baglio:2013toa}. 
The NLO EW effects were found to be sizable and affect shapes of distributions in a non-uniform way, reaching
the expected size of tens of percent in the TeV range.
The $\gamma\gamma$ channel, which is also discussed there, turns out to be quite sizable at TeV scales, in particular 
for large rapidity differences of the W~bosons.
Finally, the quark--photon channels, which were calculated in \citere{Baglio:2013toa}, can also reach 
tens of percent with respect to the LO prediction at high energies, but this enhancement concentrates on regions of phase space
where QCD corrections dominate over the LO contributions, rendering the total impact of the quark--photon channels on the
full cross section small.%
\footnote{The photonic effects of elastic proton scattering have been discussed in
\citere{Luszczak:2013ata}. In this paper, however, we deal with inelastic scattering only.}

Clearly, the next step should therefore be aimed towards the calculation of the full EW corrections for weak-boson pair production with consideration of off-shell effects and decay correlations, i.e.\ the computation of the NLO EW corrections for the processes $\Pp\Pp\to 4\,$fermions. 
In experiments, the cleanest signatures for many observables
in the $\Pp\Pp\to \PWp\PWm\to 4$~fermions analyses
are expected from the fully leptonic decays of the weak bosons. 
To avoid mixing with the $\Pp\Pp\to \PZ\PZ\to\ell^+\ell^-\nu_\ell\bar\nu_\ell$ production process, one may focus on a final state with two charged leptons of different type, and the associated neutrinos, such as $\mvev$. Providing results for $\Pp\Pp\to\mvev$ at $\mc{O}(\alpha^5)$ is the central aim of this work.

%
The calculation of the full $\mc{O}(\alpha)$ corrections to $\Pp\Pp\to\mvev$ is a formidable task. 
Although such a full NLO EW calculation is technically possible with modern computer algebra techniques, 
and in fact has been worked out for $\Pep\Pem$ annihilation some years 
ago~\cite{Denner:2005es},
it seems desirable to provide approximative results that are almost equivalent to the full EW
corrections in phenomenologically relevant applications, but much more compact and thus suitable for 
computationally intensive in-depth numerical studies. 
We therefore employ the so-called {\em double pole approximation}~(DPA) -- 
a technique that has been developed~\cite{Stuart:1991xk} 
and applied~\cite{Beenakker:1998gr,Jadach:1998tz,Denner:1999kn,Denner:2000bj}
in the context of four-fermion production processes at lepton 
colliders in different variants.
In our calculation for W-pair production at the LHC we follow the 
strategy~\cite{Denner:1999gp,Denner:1999kn,Denner:2000bj}
employed in the Monte Carlo generator {\tt RacoonWW}~\cite{Denner:2002cg}. 
Generally, in DPA
only those contributions of the NLO EW corrections to $\Pe^+ \Pe^-\to \PWp\PWm\to 4~\mr{fermions}$ are retained 
that are enhanced by two resonant weak-boson propagators and thus by a factor of $\MW/\Gamma_\PW$ compared to 
all remaining contributions. With respect to the LO results, naive power counting suggests that the 
suppression of the neglected terms in the EW corrections is typically of the order of 
$\alpha/\pi\times \Gamma_\PW/\MW$, and thus well below a phenomenologically significant level. Naturally, this way of 
argument is true only in an experimental setup that favors configurations where both weak bosons are near their
mass shell. 
A detailed comparison of DPA results of {\tt RacoonWW}~\cite{Denner:1999kn,Denner:2000bj,Denner:2002cg}
with the full NLO EW results of \citere{Denner:2005es} has revealed that 
this DPA in fact reproduces integrated cross sections from slightly above the W-pair threshold
up to a centre-of-mass energy of $500\GeV$ ($2\TeV$) within $0.5\%$ ($2\%$).

Technically, the DPA is based on a systematic expansion of matrix elements about the two
W-boson resonance poles, thereby keeping only the leading terms.
According to this prescription, any diagram without the desired resonance structure
is dropped, but the relevant resonance diagrams are decomposed into on-shell production
and on-shell decay subprocesses, linked by off-shell propagators describing the
resonances. Spin correlations between the various subprocesses are maintained in this expansion.
Since virtual corrections involve LO kinematics, this strategy
can be adopted in a straightforward manner for the virtual EW corrections to 
$\Pp\Pp\to\mvev$.
In the DPA, the full set of virtual contributions are grouped into {\it factorizable} 
corrections to only the production of two on-shell $\PW$~bosons or their decays, and 
{\it non-factorizable} corrections that connect production and decay or the two decay subprocesses
via soft-photon exchange. 
Since only soft particle exchange in the loop is relevant in the non-factorizable contributions, they can be isolated and calculated
in a straightforward way and possess a quite simple structure. 
If a DPA is also employed for the real-emission contributions, care has to be exercised, 
since photon radiation off the W~bosons leads to overlapping resonances that have to be
disentangled when attributing photon emission to production and decay subprocesses.
Such a strategy is followed in \citeres{Beenakker:1998gr,Jadach:1998tz}, where a DPA is applied to
both virtual and real EW corrections.
Instead, in \citeres{Denner:1999gp,Denner:1999kn,Denner:2000bj,Denner:2002cg} full expressions are used 
for the real contributions, and only the virtual corrections are treated in the DPA. 
In this way, the complications associated with the real-emission kinematics are circumvented.
In our work, we 
adapt the latter method to four-fermion production processes at hadron colliders
and make use of full matrix-element calculations for the real corrections.
This has the additional advantage that a possible future promotion of the calculation to a full NLO EW
calculation without DPA requires only a change in the virtual corrections.
Globally good agreement between the different variants of performing a DPA was found
for W-pair production at lepton colliders~\cite{Grunewald:2000ju}.

Actually a 
complete NLO EW calculation to $\Pp\Pp\to\PWp\PWm\to4\,$fermions would require the inclusion
of NLO EW corrections to the $\ga\ga$ channel as well. Those corrections were worked out in
\citere{Bredenstein:2004ef} in DPA for a possible future $\ga\ga$ collider
and found to be of moderate size. Since the contribution of the whole $\ga\ga$ channel is small
compared to the quark-initiated channels, we can, however, safely restrict the treatment of the
$\ga\ga$ contribution to LO.

The paper is organized as follows:
The technical details of our calculation are described in 
\refse{sec:calculation}. Phenomenological results are presented in \refse{sec:pheno}. 
Our conclusions are given in \refse{sec:concl}. 

%
%
\section{Technical details of the calculation}
\label{sec:calculation}
\subsection{Tree-level contributions and general setup}
\label{ssec:born}
At the LHC, the production of a four-fermion final state at tree level mainly proceeds via the annihilation of an antiquark--quark pair, 
\beq
\bar q q \to \mvev\,,
\eeq
with $q$ denoting a light quark, $q=\Pu,\Pd,\Pc,\Ps$, where all $q$ are taken massless.
Note that for $\bar\Pb\Pb$ collisions,
\beq
\bar \Pb \Pb \to \mvev\,,
\eeq
a (massive) top-quark appears as an intermediate state.
Owing to the suppression of the bottom-quark density the contribution of the $\bar\Pb\Pb$ channel
is already very small at LO, so that we do not include any effects of initial-state bottom-quarks
in the EW corrections.

Because of the non-vanishing photon content of the proton, 
non-negligible contributions arise from photon--photon-induced subprocesses of the type
\beq
\gamma\gamma \to \mvev\,.
\eeq
The amplitudes for both types of production mode include diagrams with two on-shell $\PW$-boson propagators, but 
also singly- and non-resonant ``background'' graphs, c.f.~\reffi{fig:lo-graphs}. 
%
\bfi
\centerline{
\begin{picture}(120,100)(0,0)
\ArrowLine(50,70)(15,70)
\ArrowLine(15,20)(50,20)
\Vertex(50,70){2}
\Vertex(50,20){2}
\ArrowLine(50,20)(50,70)
\Photon(50,70)(90,70){2}{5}
\Photon(50,20)(90,20){2}{5}
\Vertex(90,70){2}
\ArrowLine(90,70)(110,80)
\ArrowLine(110,60)(90,70)
\Vertex(90,20){2}
\ArrowLine(90,20)(110,30)
\ArrowLine(110,10)(90,20)
\put(0,70){$\bar d$}
\put(0,20){$d$}
\put(65,75){$\PW$}
\put(38,40){$u$}
\put(65,5){$\PW$}
\put(115,78){$f_1$}
\put(115,58){$\bar f_2$}
\put(115,28){$f_3$}
\put(115,8){$\bar f_4$}
\end{picture}
\hspace*{2em}
\begin{picture}(120,100)(0,0)
\ArrowLine(50,70)(15,70)
\ArrowLine(15,20)(50,20)
\Vertex(50,70){2}
\Vertex(50,20){2}
\ArrowLine(50,20)(50,70)
\Photon(50,70)(90,20){2}{7}
\Photon(50,20)(90,70){2}{7}
\Vertex(90,70){2}
\ArrowLine(90,70)(110,80)
\ArrowLine(110,60)(90,70)
\Vertex(90,20){2}
\ArrowLine(90,20)(110,30)
\ArrowLine(110,10)(90,20)
\put(0,70){$\bar u$}
\put(0,20){$u$}
\put(65,75){$\PW$}
\put(38,40){$d$}
\put(65,5){$\PW$}
\put(115,78){$f_1$}
\put(115,58){$\bar f_2$}
\put(115,28){$f_3$}
\put(115,8){$\bar f_4$}
\end{picture}
\hspace*{2em}
\begin{picture}(120,100)(0,0)
\ArrowLine(40,45)(15,70)
\ArrowLine(15,20)(40,45)
\Vertex(40,45){2}
\Vertex(70,45){2}
\Photon(40,45)(70,45){2}{4}
\Photon(70,45)(90,20){2}{4}
\Photon(70,45)(90,70){2}{4}
\Vertex(90,70){2}
\ArrowLine(90,70)(110,80)
\ArrowLine(110,60)(90,70)
\Vertex(90,20){2}
\ArrowLine(90,20)(110,30)
\ArrowLine(110,10)(90,20)
\put(0,70){$\bar q$}
\put(0,20){$q$}
\put(65,60){$\PW$}
\put(43,30){$\ga/\PZ$}
\put(65,20){$\PW$}
\put(115,78){$f_3$}
\put(115,58){$\bar f_4$}
\put(115,28){$f_1$}
\put(115,8){$\bar f_2$}
\end{picture}
}
\centerline{
\begin{picture}(120,100)(0,0)
\Photon(12,70)(45,45){2}{4}
\Photon(12,20)(45,45){2}{4}
\Vertex(45,45){2}
\Photon(45,45)(80,70){2}{4}
\Photon(45,45)(80,20){2}{4}
\Vertex(80,70){2}
\ArrowLine(80,70)(100,80)
\ArrowLine(100,60)(80,70)
\Vertex(80,20){2}
\ArrowLine(80,20)(100,30)
\ArrowLine(100,10)(80,20)
\put(0,70){$\ga$}
\put(0,20){$\ga$}
\put(52,64){$\PW$}
\put(52,18){$\PW$}
\put(105,78){$f_1$}
\put(105,58){$\bar f_2$}
\put(105,28){$f_3$}
\put(105,8){$\bar f_4$}
\end{picture}
\hspace*{2em}
\begin{picture}(120,100)(0,0)
\Photon(15,70)(50,70){2}{4}
\Photon(15,20)(50,20){2}{4}
\Vertex(50,70){2}
\Vertex(50,20){2}
\Photon(50,70)(50,20){2}{6}
\Photon(50,70)(90,70){2}{5}
\Photon(50,20)(90,20){2}{5}
\Vertex(90,70){2}
\ArrowLine(90,70)(110,80)
\ArrowLine(110,60)(90,70)
\Vertex(90,20){2}
\ArrowLine(90,20)(110,30)
\ArrowLine(110,10)(90,20)
\put(0,70){$\ga$}
\put(0,20){$\ga$}
\put(65,75){$\PW$}
\put(55,40){$\PW$}
\put(65,5){$\PW$}
\put(115,78){$f_1$}
\put(115,58){$\bar f_2$}
\put(115,28){$f_3$}
\put(115,8){$\bar f_4$}
\end{picture}
\hspace*{2em}
\begin{picture}(120,100)(0,0)
\Photon(15,70)(50,20){2}{6}
\Photon(15,20)(50,70){2}{6}
\Vertex(50,70){2}
\Vertex(50,20){2}
\Photon(50,70)(50,20){2}{6}
\Photon(50,70)(90,70){2}{5}
\Photon(50,20)(90,20){2}{5}
\Vertex(90,70){2}
\ArrowLine(90,70)(110,80)
\ArrowLine(110,60)(90,70)
\Vertex(90,20){2}
\ArrowLine(90,20)(110,30)
\ArrowLine(110,10)(90,20)
\put(0,70){$\ga$}
\put(0,20){$\ga$}
\put(65,75){$\PW$}
\put(55,40){$\PW$}
\put(65,5){$\PW$}
\put(115,78){$f_1$}
\put(115,58){$\bar f_2$}
\put(115,28){$f_3$}
\put(115,8){$\bar f_4$}
\end{picture}
}
\centerline{
\begin{picture}(160,100)(0,0)
\ArrowLine(40,45)(15,70)
\ArrowLine(15,20)(40,45)
\Vertex(40,45){2}
\Vertex(70,45){2}
\Photon(40,45)(70,45){2}{4}
\Photon(90,62.5)(120,45){2}{4}
\Vertex(90,62.5){2}
\ArrowLine(70,45)( 90,62.5)
\ArrowLine(90,62.5)(110,80)
\ArrowLine(140,60)(120,45)
\Vertex(120,45){2}
\ArrowLine(120,45)(140,30)
\ArrowLine(110,10)(70,45)
\put(0,70){$\bar q$}
\put(0,20){$q$}
\put(95,40){$\PW$}
\put(65,60){$f_2$}
\put(43,30){$\ga/\PZ$}
\put(115,78){$f_1$}
\put(145,58){$\bar f_4$}
\put(145,28){$f_3$}
\put(115,8){$\bar f_2$}
\end{picture}
\hspace*{2em}
\begin{picture}(120,100)(0,0)
\Photon(15,70)(50,70){2}{4}
\Photon(15,20)(50,20){2}{4}
\Vertex(50,70){2}
\Vertex(50,20){2}
\ArrowLine(50,20)(50,70)
\ArrowLine(50,70)(90,70)
\ArrowLine(110,10)(50,20)
\Vertex(90,70){2}
\Vertex(120,45){2}
\Photon(120,45)(90,70){2}{4}
\ArrowLine(90,70)(110,80)
\ArrowLine(120,45)(140,30)
\ArrowLine(140,60)(120,45)
\put(90,45){$\PW$}
\put(0,70){$\ga$}
\put(0,20){$\ga$}
\put(65,77){$f_2$}
\put(55,40){$f_2$}
\put(115,78){$f_1$}
\put(145,58){$\bar f_4$}
\put(145,28){$f_3$}
\put(115,8){$\bar f_2$}
\end{picture}
}
\caption{A representative set of diagrams contributing to $\Pp\Pp\to\PWp\PWm\to f_1\bar f_2 f_3\bar f_4$ at tree level:
doubly-resonant contributions of the $\bar qq$ channels (first line),
doubly-resonant contributions of the $\ga\ga$ channel (second line),
examples for ``background diagrams'' with only a single W~resonance (third line).}
\label{fig:lo-graphs}
\efi
%
We fully take into account all these contributions in our LO calculation. 
The masses of all external fermions 
are disregarded, unless they are needed to regularize infrared (IR) singularities that emerge in the calculation 
of electroweak corrections. 
Neglecting the extremely suppressed quark mixing of the third generation with the first two generations,
the remaining two-dimensional Cabibbo mixing drops out after summing over intermediate light-quark states
owing to the unitarity of the quark-mixing matrix, so that we may perform the whole calculation with
the quark mixing matrix set to the unit matrix.
The partonic cross sections 
from $\Ps$ and $\Pc$ quarks are, thus, exact copies of the $\Pd$- and $\Pu$-initiated 
processes, respectively.  

According to these considerations, we need to compute the full partonic matrix elements for the subprocesses
\beq
\bar \Pu\Pu / \bar \Pd\Pd / \bar \Pb\Pb / \ga\ga\to \mvev.
\eeq
The actual calculation of the LO matrix elements proceeds along the same lines as for the processes 
$\Pe^+ \Pe^-\to 4\,$fermions and $\ga\ga\to 4\,$fermions, outlined in \citeres{Denner:1999kn}
and \cite{Bredenstein:2004ef}, respectively.
Making use of the Weyl-van-der-Waarden spinor formalism as formulated in \citere{Dittmaier:1998nn}, 
for each subprocess compact results are obtained that are amenable for implementation in an efficient Monte 
Carlo program. 
For a gauge-invariant description of the W- and Z-boson resonances appearing in the off-shell matrix elements
we employ the complex-mass scheme as introduced in \citere{Denner:1999kn} for tree-level processes, i.e.\
we consistently use complex mass parameters $\mu_V^2=M_V^2-\ri M_V\Gamma_V$ for $V=\PW,\PZ$ and derive
all EW couplings from the complex ratio $\cw=\mu_\PW/\mu_\PZ$ in our calculation of the LO and
real-emission contributions to the cross section.
%
%
%
\subsection{Structure of the next-to-leading order contributions}
\label{se:virtstruct}
The computation of the NLO EW corrections to $\Pp\Pp\to\mvev$ in the framework 
of the DPA requires the calculation of 
virtual one-loop and real-emission contributions. 
We implement the complete virtual EW corrections to the production of a $\PWp\PWm$ pair and its decay as well as the non-factorizable corrections to $\bar qq \to\mvev$ at one loop in the DPA. For the real-emission subprocesses, $\bar qq\to \mvev\gamma$ and all crossing-related reactions, we employ full matrix elements at order $\mc{O}(\alpha^5)$. The cancellation of IR singularities that arise in both, virtual and real-emission contributions, is performed by means of the dipole subtraction approach of \citeres{Dittmaier:1999mb,Dittmaier:2008md}. The numerical impact of the photon-induced subprocesses, $\gamma\gamma\to\mvev$, is quite small already at tree level. 
As anticipated already in the introduction, 
we therefore do not include EW corrections to this class of contributions, which were found to
be of moderate size in \citere{Bredenstein:2004ef}.

The hadronic cross section for $\Pp\Pp\to \mvev$ at NLO in $\alpha$ is of the form
\beqar
\label{eq:sigma-nlo}
\sigma^\mr{NLO}_{\Pp\Pp}
& = & 
\int_0^1 \rd x_1 \int_0^1 \rd x_2\,
\Biggl\{ \Biggl(
	\sum_{q = \Pu,\Pd,\Pc,\Ps} f_{\bar q}(x_1,\muf)  f_{q}(x_2,\muf) 
\non\\
&&
\left.
	{}\times
	\left[
	\int_4 \rd\hat{\sigma}_{\bar qq}^\mr{LO}+ 
	\int_4 \rd\hat{\sigma}_{\bar qq}^\mr{virt}+ 
	\int_5 \rd\hat{\sigma}_{\bar qq}^\mr{real} +
 	\int_0^1 \rd x\int_4 \rd\hat{\sigma}_{\bar qq}^\mr{fact} 
	\right]
	{}+
	(q\leftrightarrow \bar q) \Biggr)
\right.
\non\\
&&
\left.
	{}+
	f_\gamma(x_1,\muf)\sum_{q = \Pu,\Pd,\Pc,\Ps} 
	\Biggl(
		f_{q}(x_2,\muf) 
		\left[
		\int_5 \rd\hat{\sigma}_{\gamma q}^\mr{real} + 
		\int_0^1 \rd x\int_4 \rd\hat{\sigma}_{\gamma q}^\mr{fact} 
		\right]
	+(q\leftrightarrow \bar q)
	\Biggr)
\right.
\non\\
&&
\left.
	{}+
	f_\gamma(x_2,\muf)\sum_{q = \Pu,\Pd,\Pc,\Ps} 
	\Biggl(
		f_{\bar q}(x_1,\muf) 
		\left[
		\int_5 \rd\hat{\sigma}_{\bar q\gamma}^\mr{real} + 
		\int_0^1 \rd x\int_4 \rd\hat{\sigma}_{\bar q\gamma}^\mr{fact} 
		\right]
	+(\bar q\leftrightarrow q)
	\Biggr)
\right.
\non\\
&&
\left.
	{}+
	\Biggl(f_{\bar\Pb}(x_1,\muf)f_{\Pb}(x_2,\muf)
	\int_4 \rd\hat{\sigma}_{\bar\Pb\Pb}^\mr{LO}
	+(\Pb\leftrightarrow \bar \Pb)
	\Biggr)
%
\right.
\non\\
&&
	{}+
	f_\gamma(x_1,\muf)f_\gamma(x_2,\muf)
	\int_4 \rd\hat{\sigma}_{\gamma\gamma}^\mr{LO}
\Biggr\}\,.
\eeqar
Here, the $f_a(x_i,\muf)$ 
are the parton distribution functions (PDFs) 
describing the (generalized) probability for finding a parton of type $a$ at a 
(factorization) scale $\muf$ in the proton that carries a fraction $x_i$ 
of the parent's momentum,
and the subscripts ``4'' or ``5'' on the integrals of $\rd\hat{\sigma}$ refer to the
number of final-state particles to be integrated over in phase space.
The integrals $\int\rd x$ in front of the factorization contributions $\rd\hat{\sigma}_{ab}^\mr{fact}$,
furthermore, indicate that there is an additional convolution over a variable $x$
involved that controls the energy loss by collinear particle emission from an incoming parton.

Each $\rd\hat{\sigma}_{ab}^\mr{LO}$ denotes a lowest-order partonic cross section for the subprocess $ab\to \mvev$,
\beq
\int_4\rd\hat{\sigma}_{ab}^\mr{LO} = 
F(x_1 x_2)
\int_4\rd\Phi_{4} \,\sumav \left|\mc{M}_\mr{Born}^{ab\to\,\mvev}\right|^2\,,
\eeq
with the four-particle phase-space factor 
$\rd\Phi_{4}$ and the corresponding tree-level matrix element $\mc{M}_\mr{Born}^{ab\to\mvev}$. 
Here and in the following, the symbol $\sumav$ indicates appropriate summation over spin and colour 
degrees of freedom for the outgoing and 
the corresponding average for the incoming particles. 
The flux factor,
\beq
F(x) = \frac{1}{2 x \spp}\,,
\eeq 
is defined via the fraction $x$ of the squared proton--proton centre-of-mass energy $\spp$ that is taken by the two scattering partons.

The real-emission cross sections are defined analogously as
\beq
\int_5\rd\hat{\sigma}_{ab}^\mr{real} = 
F(x_1 x_2)\int_5\rd\Phi_{5} \,\sumav\left|\mc{M}_\mr{real}^{ab\to\,\mvev c}\right|^2
\eeq
and contain the partonic matrix elements for the full tree-level processes $\bar qq\to\,\mvev\gamma$, 
$\gamma q\to \mvev q$ and likewise for $\bar q\gamma$ collisions. 
The matrix elements of all those
$2\to5$ particle processes are related via crossing, which can be exploited to technically simplify the
amplitude evaluations.
Each real-emission cross section can be decomposed into a sum of finite and singular terms,
\beq
\int_5\rd\hat{\sigma}_{ab}^\mr{real} = 
\int_5\rd\hat{\sigma}_{ab}^\mr{real,fin} + 
\int_5\rd\hat{\sigma}_{ab}^\mr{real,sing},
\eeq
where we define the finite parts $\rd\hat{\sigma}_{ab}^\mr{real,fin}$ upon subtracting the auxiliary
cross sections $\rd\hat{\sigma}_{ab}^\mr{real,sing}$ 
according to the dipole subtraction technique, as described in detail below.
This auxiliary cross section is constructed in such a way that the one-particle phase space
containing the soft and collinear IR singularity can be integrated out analytically,
\beq
\int_5\rd\hat{\sigma}_{ab}^\mr{real,sing} =
\int_0^1\rd x\,\int_4\rd\hat{\sigma}_{ab}^\mr{real,conv} +
\int_4\rd\hat{\sigma}_{ab}^\mr{real,endp},
\label{eq:realsing}
\eeq
so that the ``endpoint'' part
$\rd\hat{\sigma}_{ab}^\mr{real,endp}$ contains all real-emission IR singularities with LO kinematics
and the parameter $x$ in the ``convolution part'' $\rd\hat{\sigma}_{ab}^\mr{real,conv}$
plays the same role as in the factorization contribution $\rd\hat{\sigma}_{ab}^\mr{fact}$.
Adding $\rd\hat{\sigma}_{ab}^\mr{fact}$, 
which results from a redefinition of the PDFs, 
to $\rd\hat{\sigma}_{ab}^\mr{real,conv}$ by construction compensates all IR divergences in the convolution part,
which originate from collinear initial-state splittings.

The endpoint part $\rd\hat{\sigma}_{ab}^\mr{real,endp}$ 
can be combined with the virtual corrections, so that all IR singularities cancel in this
sum as well. Note that $\rd\hat{\sigma}_{ab}^\mr{real,endp}$ exists only for the $\bar qq$ channel.
Consequently, the finite part of the virtual corrections for the subprocess $\bar qq\to \mvev$ is defined by
\beq 
\label{eq:virt-full}
\int_4\rd\hat{\sigma}_{\bar qq}^\mr{virt,fin} =
\int_4\rd\hat{\sigma}_{\bar qq}^\mr{virt} +
\int_4\rd\hat{\sigma}_{\bar qq}^\mr{real,endp}. 
\eeq
As explained above, we want to include the finite part of the virtual corrections in the DPA.
Technically, this is achieved by computing the entire virtual cross section and the endpoint parts of the
real corrections in the framework of the DPA, thereby taking care that exactly the same systematic
pole expansions is made in either part, in order not to spoil the cancellation of IR divergences.
The DPA splits the full virtual corrections into {\it factorizable} and {\it non-factorizable}
contributions. The former, by definition, 
consist of the corrections to either on-shell W~production 
or on-shell W~decay, the latter contain the remaining doubly-resonant corrections which are
entirely due to soft-photon exchange between production and decays or between the two different decay processes.
In the non-factorizable corrections, the off-shell propagators of the two W~bosons are intertwined in a 
complicated way, but the whole correction takes the form of a correction factor $\delta^\mr{virt}_\mr{nfact}$
to the Born cross section. The virtual correction in DPA, thus, can be written as
\beqar
\int_4\rd\hat{\sigma}_{\bar qq}^\mr{virt,DPA}
&=&
F(x_1 x_2)
\int_4\rd\Phi_{4}
\left\{
   2\,\mr{Re}\left[
	 	\left(
		\mc{M}_\mr{Born,DPA}^{\bar qq\to \,\PW\PW\to\,\mvev}
		\right)^\star
	 	\delta\mc{M}_\mr{virt,fact,DPA}^{\bar qq\to\, \PW\PW\to\,\mvev}
	     \right]
\right.	
	\non\\
	&&
\left.
   + \left|\mc{M}_\mr{Born,DPA}^{\bar qq\to \,\PW\PW\to \,\mvev}\right|^2
     \delta^\mr{virt}_\mr{nfact}
\right\}\,.
\label{eq:virtDPA}
\eeqar		
Here, $\delta\mc{M}_\mr{virt,fact,DPA}^{\bar qq\to\, \PW\PW\to\,\mvev}$ denotes the matrix element for the factorizable virtual corrections, and $\delta^\mr{virt}_\mr{nfact}$ stems from the non-factorizable virtual contributions in the DPA.

Schematically, the various cross-section contributions to a quark-initiated subprocess $\bar qq\to\mvev$ can then be summarized as
\beqar
\int \rd\hat{\sigma}_{\bar q q} = 
\int_4 \rd\hat{\sigma}_{\bar q q}^\mr{LO}
+\int_4 \rd\hat{\sigma}_{\bar q q}^\mr{virt,fin,DPA}
+\int_5 \rd\hat{\sigma}_{\bar q q}^\mr{real,fin}
+\int_0^1\rd x\int_4 \left(\rd\hat{\sigma}_{\bar q q}^\mr{fact}+\rd\hat{\sigma}_{\bar q q}^\mr{real,conv}\right)\,.
\eeqar
An analogous expression is obtained for the $q\bar q$ initial states. 
For the photon-induced processes, only real-emission and factorization contributions appear,
\beqar
\int \rd\hat{\sigma}_{q \gamma} = 
\int_5 \rd\hat{\sigma}_{q \gamma}^\mr{real,fin}+
\int_0^1\rd x\int_4 \left(\rd\hat{\sigma}_{q \gamma}^\mr{fact}+\rd\hat{\sigma}_{q\gamma}^\mr{real,conv}\right)\,,
\eeqar
and analogously for antiquarks.
Here, all IR singularities associated with the $2\to 5$ scattering process are factorized 
into the PDFs of the proton.

%
\subsection{Virtual corrections}
\label{ssec:virtuals}
As explained in some detail in \citere{Denner:2000bj}, the one-loop corrections to four-fermion production 
processes in the framework of the DPA are classified in terms of factorizable and non-factorizable contributions. 
Here we provide some details concerning their explicit calculation.
Throughout, in reactions of the type
\beq
\bar q(q_1)\, q(q_2) \to
\nu_\mu(q_3)\, \mu^+(q_4)\, e^-(q_5)\, \bar\nu_e(q_6)
\label{eq:process}
\eeq
we generically denote the quark and lepton momenta by $q_i$ with $i=1,\dots,6$.
The electric charge and mass of the fermion with momentum $q_i$ are denoted 
by $Q_i$ and $m_i$,
respectively. 

\paragraph{(i) Factorizable corrections}

The {\em factorizable contributions} are obtained by selecting all one-loop diagrams that contain two resonant 
W-boson propagators and loop corrections either associated with the production of two on-shell W-bosons 
or the decay of one of these weak bosons. The relevant generic diagrammatic structure is shown in \reffi{fig:factdiag}.
\begin{figure}
\centerline{
\begin{picture}(200,105)(0,0)
\ArrowLine(30,50)( 5, 95)
\ArrowLine( 5, 5)(30, 50)
\Photon(30,50)(150,80){2}{11}
\Photon(30,50)(150,20){2}{11}
\ArrowLine(150,80)(190, 95)
\ArrowLine(190,65)(150,80)
\ArrowLine(190, 5)(150,20)
\ArrowLine(150,20)(190,35)
\GCirc(30,50){10}{.5}
\GCirc(90,65){10}{1}
\GCirc(90,35){10}{1}
\GCirc(150,80){10}{.5}
\GCirc(150,20){10}{.5}
\DashLine( 70,0)( 70,100){2}
\DashLine(110,0)(110,100){2}
\put(50,26){W}
\put(50,68){W}
\put(115,13){W}
\put(115,82){W}
\put(-12, 0){$q$}
\put(-12,95){$\bar q$}
\put(195, 1){$\bar f_4$}
\put(195,34){$f_3$}
\put(195,60){$\bar f_2$}
\put(195,95){$f_1$}
\put(-25,-15){\footnotesize On-shell production}
\put(120,-15){\footnotesize On-shell decays}
\end{picture}
}
\vspace*{1em}
\caption{Generic diagram for virtual factorizable corrections to
$\bar qq\to\, \PW\PW\to4\,$fermions.}
\label{fig:factdiag}
\efi
The physical polarization degrees of freedom of the W bosons 
that connect the production with the decay processes, $\lambda_\pm$, have to be summed over coherently.  
In order to maintain gauge invariance, apart from the two resonant propagator factors the entire amplitude 
is evaluated on shell, i.e.\ the momenta of the $\PW^+$ and $\PW^-$ bosons, 
\beq
k_+ = q_3+q_4, \qquad k_- = q_5+q_6,
\label{eq:Wmom} 
\eeq
are globally replaced with their on-shell projections  
$\hat{k}_+$ and $\hat{k}_-$, which fulfill $\hat{k}_+^2=\hat{k}_-^2=\MW^2$. 
To this end, each phase-space point with general off-shell kinematics has to be identified with a 
phase-space point for on-shell W bosons such that $|k_\pm^\mu-\hat k_\pm^\mu|^2$ is of order
$\ord(k_\pm^2-\MW^2)$ for each component $\mu$. Different variants of the actual projection are possible and 
equally viable leading to results that differ only by effects beyond DPA accuracy. 
We stick to the version of \citere{Denner:2000bj} (see App.~A in there). 
Throughout, we denote momenta associated with on-shell 
kinematics by hats.

The entire contribution of the factorizable corrections to the virtual amplitude can then be written in the form
\beqar
\delta\mc{M}_\mr{virt,fact,DPA}^{\bar qq\to \PWp\PWm\to\mvev}
&=&
\sum_{\lambda_+,\lambda_-}
\frac{1}{
\left[
k_+^2-\MW^2+\ri\MW\Gamma_\PW
\right]
\left[
k_-^2-\MW^2+\ri\MW\Gamma_\PW
\right]
}
\non\\
&& {}\times
\left\{
\delta\mc{M}^{\bar qq\to \PWp\PWm}
\mc{M}^{W^+\to\nu_\mu \mu^+}_\mr{Born}
\mc{M}^{W^-\to \Pe^-\bar\nu_\Pe}_\mr{Born}
\right.
\non\\
&&
\left.
\quad {}+
\mc{M}_\mr{Born}^{\bar qq\to \PWp\PWm}
\delta\mc{M}^{W^+\to\nu_\mu \mu^+}
\mc{M}^{W^-\to \Pe^-\bar\nu_\Pe}_\mr{Born}
\right.
\non\\
&&
\left.
\quad {}+
\mc{M}_\mr{Born}^{\bar qq\to \PWp\PWm}
\mc{M}^{W^+\to\nu_\mu \mu^+}_\mr{Born}
\delta\mc{M}^{W^-\to \Pe^-\bar\nu_\Pe}
\right\}\,,
\eeqar
with the summation extending over the helicities $\lambda_\pm$ of the $\PW^\pm$~bosons. 
All terms within the curly brackets are understood to be evaluated for on-shell kinematics. The $\delta\mc{M}$ denote one-loop contributions. 

The actual calculation of the $\delta\mc{M}_\mr{virt,fact,DPA}^{\bar qq\to \PWp\PWm\to\mvev}$ proceeds along the 
lines of \citere{Denner:2000bj} for the related case of four-fermion production in $\Pe^+\Pe^-$ collisions. 
The entire amplitude is expressed in terms of so-called standard matrix elements (SMEs), $\mc{M}_n^\sigma$, which contain the spin structure of the external particles for specific helicity combinations, and appropriate coefficient functions, $F_n^\sigma$. The latter only depend on the Mandelstam invariants,  
\beqar
\hat{s} &=& (q_1 + q_2)^2 = (\hat{k}_+ + \hat{k}_-)^2\,,
\non\\
\hat{t}&=&  (q_1 - \hat{k}_+)^2= (q_2 - \hat{k}_-)^2
= \MW^2-\frac{\hat s}{2}
\left(1-\beta\cos\theta\right)
\,,
\eeqar
where the $q_{1,2}$ are the momenta of the incoming partons,  $\beta = \sqrt{1-4 \MW^2/\hat s}$, and $\theta$ denotes the scattering angle of the outgoing W bosons with respect to the beam axis in the partonic centre-of-mass frame.
The SMEs for the on-shell production of a W~pair produced via the specific quark chirality $\sigma=\pm 1$ are contracted with off-shell $\PW$ propagators and tree-level decay matrix elements that are also calculated for on-shell kinematics. 
Explicitly, one finds
\beq
\delta\mc{M}_\mr{virt,fact,DPA}^{\bar qq\to \PWp\PWm\to\mvev,\sigma}
(q_1,q_2,\{q_i\},k_+^2,k_-^2) =
\sum_{n=1}^7
F_n^\sigma(\hat{s},\hat{t})
\mc{M}_n^{\sigma}
(q_1,q_2,\{\hat q_i\},k_+^2,k_-^2)\,,
\label{eq:Mvirt}
\eeq
where $\{q_i\}$ stands for the set of outgoing lepton momenta.
Note that the SMEs are just tree-level expressions; their precise definition is given
in Eq.~(3.1) of \citere{Denner:2000bj}.
All information on the loop structure of the reaction is contained in the form factors, 
$F_n^\sigma(\hat{s},\hat{t})$, which besides the incoming partons' momenta only depend on the on-shell-projected 
momenta of the W~bosons, 
but not on the momenta of their decay products. The actual calculation of the $F_n^\sigma$ proceeds via standard methods of one-loop
calculations applied to the $2\to2$ particle production and $1\to2$ particle decay subprocesses.
Since the one-loop corrections to these subprocesses do not involve unstable-particle effects, the traditional EW on-shell
renormalization with real mass parameters, as described for instance in \citere{Denner:1991kt}, can be applied without modification.
In detail, after 
generating the respective one-loop amplitudes with the program 
{\tt FeynArts}~\cite{Kublbeck:1990xc}, we algebraically reduce the amplitudes to the standard form
\refeq{eq:Mvirt}, once using the program {\tt FormCalc}~\cite{Hahn:1998yk}
and alternatively inhouse {\tt Mathematica} routines.
The respective automatically generated {\tt Fortran} output is evaluated with the help
of either of the numerical loop libraries {\tt LoopTools}~\cite{Hahn:1998yk} or 
{\tt Collier}, where the latter is an implementation of one-loop tensor and scalar integrals
described in \citeres{Passarino:1979jh,'tHooft:1979xw,Dittmaier:2003bc}.

Finally, we recall that the coefficient functions $F_n^\sigma(\hat{s},\hat{t})$, which depend only
on the scattering energy $\sqrt{\hat s}$ and the scattering angle $\theta$, can be very
efficiently evaluated via a series expansion in terms of a Legendre polynomials
$P_l(\cos\theta)$. Details of such an approach can be found in 
Sect.~3.1.2 of \citere{Denner:2000bj} and Sect.~3.2.2 of the second paper of 
\citere{Bredenstein:2004ef}.
Calculating the corresponding coefficients before the actual phase-space
integration reduces the CPU time for the evaluation of the loop diagrams during the
phase-space integration to a nearly negligible level.

\paragraph{(ii) Non-factorizable corrections}

All loop diagrams to $\bar qq\to4\,$fermions that cannot be factorized into two leading-order,
resonant W~propagators and some remaining part potentially contribute to the 
{\em non-factorizable virtual corrections}. 
Among these manifestly non-factorizable diagrams only those are relevant that lead to
doubly-resonant contributions in the on-shell limit of the two fermion--antifermion pairs
corresponding to the two decaying W~bosons. This can only happen in diagrams where a photon is exchanged
between the W-pair production process and one of the W~decay processes, or between the two W~decay processes,
as shown in \citeres{Fadin:1993dz,Melnikov:1995fx,Denner:1997ia} by an appropriate power counting
for the two singularities. If a massive particle, such as a Z~boson, is exchanged between production and decay subdiagrams,
at least one of the two decay fermion--antifermion pairs is shifted out of resonance; only for
soft-photon exchange a double resonance can occur, since a soft photon does not change the particle
kinematics of the underlying diagram.
Apart from those manifestly non-factorizable diagrams,
also some diagrams appearing already in the factorizable corrections contain non-factorizable contributions
in which the process of setting the W-boson momentum on its mass shell creates a soft singularity.
This happens in diagrams where a photon couples to an internal W~boson and an external fermion or again to 
an internal W~boson.
The difference between the full diagram and its mass-shell-projected version, which is part of the
factorizable corrections, is then part of the non-factorizable corrections. A survey of
representative diagrams contributing to the non-factorizable corrections is shown in
\reffi{fig:topo-nonf}.
\bfi
\begin{center}
{\unitlength .8pt \small
\SetScale{.8}
\begin{picture}(360,510)(0,10)
\Text(0,495)[lb]{(a) type (\mfp)}
\put(20,390){
\begin{picture}(150,100)(0,0)
\ArrowLine(30,50)( 5, 95)
\ArrowLine( 5, 5)(30, 50)
\Photon(30,50)(90,20){2}{6}
\Photon(30,50)(90,80){-2}{6}
\Vertex(60,65){2.0}
\GCirc(30,50){10}{.5}
\Vertex(90,80){2.0}
\Vertex(90,20){2.0}
\ArrowLine(90,80)(120, 95)
\ArrowLine(120,65)(90,80)
\ArrowLine(120, 5)( 90,20)
\ArrowLine( 90,20)(105,27.5)
\ArrowLine(105,27.5)(120,35)
\Vertex(105,27.5){2.0}
\Photon(60,65)(105,27.5){-2}{5}
\put(86,50){$\gamma$}
\put(63,78){$\PW$}
\put(40,65){$\PW$}
\put(52,18){$\PW$}
\put(-28, 5){$q(q_2)$}
\put(-28,90){$\bar q(q_1)$}
\put(128,90){$f_1(q_3)$}
\put(128,65){$\bar f_2(q_4)$}
\put(128,30){$f_3(q_5)$}
\put(128, 5){$\bar f_4(q_6)$}
\end{picture}
}
\Text(210,495)[lb]{(b) type (\ffp)}
\put(230,390){
\begin{picture}(120,100)(0,0)
\ArrowLine(30,50)( 5, 95)
\ArrowLine( 5, 5)(30, 50)
\Photon(30,50)(90,80){-2}{6}
\Photon(30,50)(90,20){2}{6}
\GCirc(30,50){10}{.5}
\Vertex(90,80){2.0}
\Vertex(90,20){2.0}
\ArrowLine(90,80)(120, 95)
\ArrowLine(120,65)(105,72.5)
\ArrowLine(105,72.5)(90,80)
\Vertex(105,72.5){2.0}
\ArrowLine(120, 5)( 90,20)
\ArrowLine( 90,20)(105,27.5)
\ArrowLine(105,27.5)(120,35)
\Vertex(105,27.5){2.0}
\Photon(105,27.5)(105,72.5){2}{4.5}
\put(93,47){$\gamma$}
\put(55,73){$\PW$}
\put(55,16){$\PW$}
\end{picture}
}
\Text(0,365)[lb]{(c) type (if)}
\put(20,260){
\begin{picture}(120,100)(0,0)
\ArrowLine(27,55)(15, 75)
\Vertex(15,75){2.0}
\ArrowLine(15,75)( 3, 95)
\ArrowLine( 3, 5)(30, 50)
\Photon(30,50)(90,80){-2}{6}
\Photon(30,50)(90,20){2}{6}
\GCirc(30,50){10}{.5}
\Vertex(90,80){2.0}
\Vertex(90,20){2.0}
\ArrowLine(90,80)(105,87.5)
\ArrowLine(105,87.5)(120, 95)
\ArrowLine(120,65)(90,80)
\ArrowLine(120, 5)( 90,20)
\ArrowLine( 90,20)(120,35)
\Vertex(105,87.5){2.0}
\PhotonArc(66.25,36.25)(64.25,52.9,142.9){2}{8}
\put(55,90){$\gamma$}
\put(68,55){$\PW$}
\put(55,16){$\PW$}
\end{picture}
}
\Text(210,365)[lb]{(d) type (\mmp)}
\put(230,260){
\begin{picture}(120,100)(0,0)
\ArrowLine(30,50)( 5, 95)
\ArrowLine( 5, 5)(30, 50)
\Photon(30,50)(90,80){-2}{6}
\Photon(30,50)(90,20){2}{6}
\Photon(70,30)(70,70){2}{3.5}
\Vertex(70,30){2.0}
\Vertex(70,70){2.0}
\GCirc(30,50){10}{.5}
\Vertex(90,80){2.0}
\Vertex(90,20){2.0}
\ArrowLine(90,80)(120, 95)
\ArrowLine(120,65)(90,80)
\ArrowLine(120, 5)( 90,20)
\ArrowLine( 90,20)(120,35)
\put(76,47){$\gamma$}
\put(45,68){$\PW$}
\put(45,22){$\PW$}
\put(72,83){$\PW$}
\put(72,11){$\PW$}
\end{picture}
}
\Text(0,235)[lb]{(e) type (im)}
\put(20,130){
\begin{picture}(150,100)(0,0)
\ArrowLine(27,55)(15, 75)
\Vertex(15,75){2.0}
\ArrowLine(15,75)( 3, 95)
\ArrowLine( 3, 5)(30, 50)
\Photon(30,50)(90,20){2}{6}
\Photon(30,50)(90,80){-2}{6}
\Vertex(60,65){2.0}
\GCirc(30,50){10}{.5}
\Vertex(90,80){2.0}
\Vertex(90,20){2.0}
\ArrowLine(90,80)(120, 95)
\ArrowLine(120,65)(90,80)
\ArrowLine(120, 5)( 90,20)
\ArrowLine( 90,20)(120,35)
\PhotonArc(32.5,47.5)(32.596,32.47,122.47){2}{4.5}
\put(36,92){$\gamma$}
\put(75,61){$\PW$}
\put(51,48){$\PW$}
\put(52,18){$\PW$}
\end{picture}
}
\Text(210,235)[lb]{(f) type (mf)}
\put(230,130){
\begin{picture}(120,100)(0,0)
\ArrowLine(30,50)( 3, 95)
\ArrowLine( 3, 5)(30, 50)
\Photon(30,50)(90,80){-2}{6}
\Photon(30,50)(90,20){2}{6}
\Vertex(70,70){2.0}
\GCirc(30,50){10}{.5}
\Vertex(90,80){2.0}
\Vertex(90,20){2.0}
\ArrowLine(90,80)(105,87.5)
\Vertex(105,87.5){2.0}
\ArrowLine(105,87.5)(120, 95)
\ArrowLine(120,65)(90,80)
\ArrowLine(120, 5)(90,20)
\ArrowLine(90,20)(120,35)
\PhotonArc(87.5,78.75)(19.566,26.565,206.565){2}{6}
\put(57,86){$\gamma$}
\put(77,62){$\PW$}
\put(50,48){$\PW$}
\put(55,16){$\PW$}
\end{picture}
}
\Text(0,105)[lb]{(g) type (mm)}
\put(20,0){
\begin{picture}(120,100)(0,0)
\ArrowLine(30,50)( 5, 95)
\ArrowLine( 5, 5)(30, 50)
\Photon(30,50)(90,80){-2}{6}
\Photon(30,50)(90,20){2}{6}
\Vertex(75,72.5){2.0}
\Vertex(50,60){2.0}
\GCirc(30,50){10}{.5}
\Vertex(90,80){2.0}
\Vertex(90,20){2.0}
\ArrowLine(90,80)(120, 95)
\ArrowLine(120,65)(90,80)
\ArrowLine(120, 5)( 90,20)
\ArrowLine( 90,20)(120,35)
\PhotonArc(62.5,66.25)(13.975,26.565,206.565){-2}{3.5}
\put(55,90){$\gamma$}
\put(44,45){$\PW$}
\put(62,54){$\PW$}
\put(82,64){$\PW$}
\put(55,16){$\PW$}
\end{picture}
}
\end{picture}
}
\end{center}
\caption{Classification of diagram types contributing to the
virtual non-factorizable corrections, following \citere{Denner:2000bj}.}
\label{fig:topo-nonf}
\efi

Non-factorizable EW corrections to pair production processes have first been considered in
\citere{Fadin:1993dz},
where it was shown that the sum of virtual and real non-factorizable corrections vanish
if the virtualities of the two resonances are integrated over, such as in the total cross section.
Later the sum of virtual and real non-factorizable corrections was evaluated analytically and
numerically in \citeres{Melnikov:1995fx,Denner:1997ia},
revealing that their contribution to the NLO EW corrections for W- and Z-boson pair production
processes in $\Pep\Pem$ annihilation is quite small in differential cross sections as well.
Thus, if the DPA is used for virtual {\it and} real corrections, non-factorizable corrections might
be disregarded in a reasonable approximation in many cases. However, since we have decided to
base our evaluation of real corrections on full matrix elements, we have to include the
virtual non-factorizable corrections for consistency. Otherwise, there would be a mismatch
between the IR singularities of the virtual and real corrections.

\begin{sloppypar}
As mentioned above,
we define the non-factorizable doubly-resonant contributions by subtracting the factorizable 
doubly-resonant contributions from the complete one-loop corrections to $\bar qq\to4\,$fermions 
and subsequently taking the limit of on-shell W~momenta whenever possible.
This procedure avoids double-counting and
ensures gauge invariance of the two contributions furnishing our DPA.
The diagrams contributing to this class of virtual corrections contain a photon propagator in the loop, 
and involve logarithms of the form $\ln\left(k_\pm^2-\MW^2+\ri \MW\Gamma_\PW\right)$, 
corresponding to the artificially created soft IR divergences in the factorizable corrections,
as mentioned above.
These have to be kept exactly. As a consequence of the fact that non-factorizable corrections entirely are caused
by soft photons (of energies $\lsim\Gamma_\PW$), it turns out that they take the form
of a correction factor to the LO cross section, 
as already anticipated in
Eq.~\refeq{eq:virtDPA}. Making use of the explicit results of \citere{Denner:2000bj},
this factor can be written as
\beqar
\delta_\mr{nfact}^\mr{virt} &=& 
\sum_{a=3,4} \, \sum_{b=5,6} \, (-1)^{a+b+1} \, Q_a Q_b \,
\nn\\
&& \qquad {}\times
\frac{\alpha}{\pi} \Re\biggl[
\Delta^\mr{virt}_\mr{mm} + \Delta^\mr{virt}_\mr{mm'}
-Q_q \left(\Delta^\mr{virt}_\mr{im}+\Delta^\mr{virt}_\mr{if}\right)
+ 
\Delta^\mr{virt}_\mr{mf}+\Delta^\mr{virt}_\mr{mf'}+\Delta^\mr{virt}_\mr{ff'} 
\biggr],
\eeqar 
where $Q_q$ 
is the electric charge of the incoming quark
and $Q_i$ the one of the fermion with momentum $q_i$ as specified in \refeq{eq:process}.
The explicit form of the seven contributions 
$\Delta^\mr{virt}_{\dots}(q_1,q_2;k_+,q_a;k_-,q_b)$ to the corresponding graph
types of \reffi{fig:topo-nonf} can be found in Eqs.~(3.11)--(3.17) of \citere{Denner:2000bj}, 
where the arguments $(p_+,p_-;k_+,k_2;k_-,k_3)$ of the $\Delta^\mr{virt}_{\dots}$ functions given there 
in our conventions read $(q_1,q_2;k_+,q_4;k_-,q_5)$, etc.~.
In \citere{Denner:2000bj}, 
each of these $\Delta^\mr{virt}_{\dots}$ is expressed in terms of scalar loop integrals with up to five 
denominator factors, $B_0$, $C_0$, $D_0$, $E_0$. Infrared singularities related to the photon propagator 
are regularized by a small photon mass $\lambda$. 
Whenever possible, the limits $k_\pm^2\to \MW^2$ and $\Gamma_\PW\to 0$ are taken. 
Some of the necessary integrals are given in App.~C of \citere{Denner:2000bj}, the remaining ones
can be found in \citere{Dittmaier:2003bc} and \citere{Denner:1997ia}.
\end{sloppypar}

\paragraph{(iii) Singular virtual corrections and subtraction endpoint}

As explained in \refse{se:virtstruct}, the (soft and/or collinear) IR singularities of
the virtual EW corrections exactly cancel against their counterpart in the real
photonic bremsstrahlung corrections, which is precisely contained in the ``endpoint contribution''
$\rd\hat{\sigma}_{\bar qq}^\mr{real,endp}$ of the dipole subtraction functions.
Exploiting this fact we have defined the singular virtual corrections to the
$\bar qq$ channel as
$\rd\hat{\sigma}_{\bar qq}^\mr{virt,sing}=-\rd\hat{\sigma}_{\bar qq}^\mr{real,endp}$,
leading to the definition \refeq{eq:virt-full}
of the finite virtual corrections.
Moreover, we recall that we treat only the finite virtual corrections 
$\rd\hat{\sigma}_{\bar qq}^\mr{virt,fin}$ in DPA, so 
that it is appropriate to describe $\rd\hat{\sigma}_{\bar qq}^\mr{real,endp}$ here.
In the dipole subtraction approach~\cite{Dittmaier:1999mb,Dittmaier:2008md} this contribution
is given by
\beq
\label{eq:real-endp}
\rd\hat{\sigma}_{\bar qq}^\mr{real,endp} =
-\rd\hat{\sigma}_{\bar qq}^\mr{Born}\,
\frac{\alpha}{2\pi}
\sum_{i=1}^6
\sum_{j=i+1}^6
(-1)^{i+j}
Q_i Q_j
\left[
\mc{L}(s_{ij},m_i^2)+
\mc{L}(s_{ij},m_j^2)
+C_{ij}+C_{ji}
\right]\,,
\eeq
where the $Q_i$ and $m_i$ denote charge and mass of particle $i$, respectively, and $s_{ij}=2q_i \cdot q_j$.
The function $\mc{L}$ reads
\beqar
\label{eq:l-functions}
\mc{L}(s,m^2) 
=
\ln
\left(
\frac{m^2}{s}
\right)
\ln\left(
\frac{\lambda^2}{s}
\right)
+
\ln\left(
\frac{\lambda^2}{s}
\right)
-
\frac{1}{2}
\ln^2\left(
\frac{m^2}{s}
\right)
+\frac{1}{2}
\ln\left(
\frac{m^2}{s}
\right)
\eeqar
in mass regularization, i.e.\
if soft-photonic singularities are regularized by an infinitesimal photon mass $\lambda$
and by small fermion masses $m_i$ with $m_i^2\ll\MW^2,\hat s,|\hat t|,s_{ij},\dots$.
The constants $C_{ij}$ are given by
\beqar
C_{ab} = -\frac{\pi^2}{3}+2\,,\quad
C_{ak} =  \frac{\pi^2}{6}-1\,,\quad
C_{ka} = -\frac{\pi^2}{2}+\frac{3}{2}\,,\quad
C_{kl} = -\frac{\pi^2}{3}+\frac{3}{2}\,,
\label{eq:Cij}
\eeqar
for $a,b=1,2$ and $k,l=3,4,5,6$. 
The respective expressions for the endpoint contributions needed in the DPA version of
Eq.~(\ref{eq:virt-full}) can be obtained from $d\hat{\sigma}_{ab}^\mr{real,endp}$ 
by simply replacing the full Born cross section with the corresponding DPA expression and replacing
the scalar products $s_{ij}$ by their counterparts $\hat s_{ij}$ with on-shell-projected momenta,
\beqar
\rd\hat{\sigma}_{ab}^\mr{real,endp,DPA}
&=&
-\rd\hat{\sigma}_{ab}^\mr{Born,DPA}\,
\frac{\alpha}{2\pi}
\sum_{i=1}^6
\sum_{j=i+1}^6
(-1)^{i+j}
Q_i Q_j
\left[
\mc{L}(\hat s_{ij},m_i^2)+
\mc{L}(\hat s_{ij},m_j^2)
+C_{ij}+C_{ji}
\right]\,.
\non\\
\eeqar
Finally, we note that we have performed the actual evaluation of the virtual and
real-endpoint contributions alternatively in dimensional regularization with light fermion masses
and the photon mass strictly zero, finding perfect agreement 
on the sum of virtual and real endpoint corrections with the result obtained in mass regularization.

%
\subsection{Real-emission and subtraction contributions}
\label{ssec:reals}
The $\mc{O}(\alpha)$ real-emission corrections to $\Pp\Pp\to \mvev$ comprise contributions from subprocesses of the type $\bar qq \to \mvev\gamma$ and  crossing-related reactions with a photon in the initial state, such as  
$\gamma q\to \mvev q$ and $\bar q\gamma \to \mvev\bar q$. For each of these subprocesses, the complete tree-level matrix elements are taken into account, and the occurring gauge-boson resonances are again treated in the complex-mass scheme.

For inclusive observables the real-emission contributions exhibit (soft and/or collinear) IR 
singularities that eventually are cancelled by respective divergences of the virtual corrections and factorization contributions. In the framework of a Monte Carlo program this cancellation is conveniently accomplished by means of a dipole subtraction approach. The basic idea of a subtraction formalism is to introduce an auxiliary function that matches the singularity structure of the real-emission contributions. After subtracting this function from the real-emission terms, the remainder can be integrated safely by standard numerical methods without running into singular configurations. In turn, the correction term is integrated over the singular regions analytically and serves to cancel IR singularities of the virtual contributions. 
In this work, we employ the dipole subtraction approach for processes with photon radiation off fermions as
described in \citere{Dittmaier:1999mb}. This variant assumes that we treat outgoing fermions and nearly collinear
photons as one quasi-particle upon applying an appropriate photon recombination, analogously to the 
application of a jet algorithm in QCD. Experimentally this situation is, e.g., realized in the
concept of ``dressed leptons'' in the analysis of ATLAS data (see e.g.\ \citere{Aad:2011gj}).
A generalization of the dipole subtraction approach to non-collinear-safe observables,
which is applicable to any situation without or with only partial photon recombination, is described in
\citere{Dittmaier:2008md}. The corresponding extension of our calculation with this method is left
to a forthcoming publication. 

\subsubsection{Subprocesses with final-state photons}
For subprocesses with an antiquark--quark pair in the initial state the subtraction terms
\beqar
\left|\mc{M}_{\mr{sub},ij}(\Phi_5)\right|^2
=
-(-1)^{i+j}Q_i Q_j e^2
g_{ij}(q_i,q_j,k)
\left|\mc{M}^{\bar qq\to\mvev}_{\mr{Born}}(\tilde{\Phi}_{4,ij})\right|^2\,,
\eeqar
are used, 
where $k$ denotes the momentum of the emitted photon, whereas $i$ and $j\ne i$ ($i,j=1,\ldots,6$) label the so-called {\em emitter} and {\em spectator} fermions, respectively, which determine the kinematics of a singular configuration. 
The dipole splitting functions $g_{ij}=g_{ij,+}+g_{ij,-}$, which correspond to the situation of unpolarized fermions, 
can be obtained from the polarized splitting functions $g_{ij,\pm}$ given in \citere{Dittmaier:1999mb} in a
straightforward way.
The arguments of the matrix elements $\mc{M}_{\mr{sub},ij}$ and $\mc{M}_{\mr{Born}}$, $\Phi_5$ and $\tilde{\Phi}_{4,ij}$, indicate that the respective expressions are to be evaluated for the full $2\to 5$ kinematics and for a 4-body configuration that is obtained for a specific emitter--spectator pair by an appropriate mapping of the full real-emission kinematics, respectively. 
The radiator functions $g_{ij}$ are constructed in such a way that they capture the soft and collinear 
singularities of the full real-emission contributions. Their explicit form is slightly different for initial- 
and final-state emitters and spectators. Following \citere{Dittmaier:1999mb}, for final-state emitters and 
spectators ($i,j=3,\ldots,6$) we are using
\beqar
%
g_{ij}(q_i,q_j,k) &=& \frac{2}{s_{i\ga}(1-y_{ij})}
\left[
\frac{2}{1-z_{ij}(1-y_{ij})}
-1-z_{ij}
\right]\,,
\eeqar
with
\beqar
y_{ij} &=& \frac{s_{i\ga}}{s_{ij}+s_{i\ga}+s_{j\ga}}\,,
\quad
z_{ij} = \frac{s_{ij}}{s_{ij}+s_{j\ga}}\,,
\eeqar
where we again define $s_{ij}=2q_i\cdot q_j$ and $s_{i\ga}=2q_i\cdot k$.
For a final-state emitter ($i=3,\ldots,6$) and an initial-state spectator ($j=1,2$) we have 
\beqar
%
g_{ij}(q_i,q_j,k) &=& 
\frac{2}{s_{i\ga}x_{ij}}
\left[
\frac{2}{2-x_{ij}-z_{ij}}
-1-z_{ij}
\right]\,,
\eeqar
with
\beqar
\label{eq:fin-in-xz}
x_{ij} &=&
\frac{s_{ij}+s_{j\ga}-s_{i\ga}}{s_{ij}+s_{j\ga}}\,,
\quad
z_{ij} = \frac{s_{ij}}{s_{ij}+s_{j\ga}}\,.
\eeqar
For an initial-state emitter ($i=1,2$) and a final-state spectator ($j=3,\ldots,6$) the radiator functions are of the form 
\beqar
%
g_{ij}(q_i,q_j,k) &=& 
\frac{2}{s_{i\ga}x_{ij}}
\left[
\frac{2}{2-x_{ij}-z_{ij}}
-1-x_{ij}
\right]\,,
\eeqar
with
\beqar
x_{ij} &=&
\frac{s_{ij}+s_{i\ga}-s_{j\ga}}{s_{ij}+s_{i\ga}}\,,
\quad
z_{ij} = \frac{s_{ij}}{s_{ij}+s_{i\ga}}\,,
\eeqar
and for initial-state emitters ($i=1,2$) and spectators ($j=1,2$) the $g_{ij}$ are given by  
\beqar
%
g_{ij}(q_i,q_j,k) &=& 
\frac{2}{s_{i\ga}x_{ij}}
\left[
\frac{2}{1-x_{ij}}
-1-x_{ij}
\right]\,,
\eeqar
with
\beqar
x_{ij} &=&
\frac{s_{ij}-s_{i\ga}-s_{j\ga}}{s_{ij}}\,,
\quad
y_{ij} = \frac{s_{i\ga}}{s_{ij}}\,.
\eeqar

The finite remainder of the real-emission contribution for the $\bar qq$-initiated subprocesses is obtained by subtracting the sum of all counterterms from the full $\bar qq\to \mvev\gamma$ cross section,
\beqar
\int_5
\rd\hat{\sigma}_{\bar qq}^\mr{real,fin} 
&=&
F(x_1 x_2)
\int_5\rd\Phi_5 \left[
\sumav
\left|\mc{M}^{\bar qq\to\mvev\gamma}_{\mr{real}}\right|^2\Theta(\Phi_5)
-
\sum_{i,j=1 \atop i\neq j}^6
\sumav
\left|\mc{M}_{\mr{sub},ij}\right|^2\Theta(\tilde{\Phi}_{4,ij})
\right]\,.
\non\\
\eeqar
Here, the $\Theta$ functions can assume the values 0 and 1. They describe the effect of separation cuts and a 
possible photon recombination procedure which ensures that soft or collinear photons are combined with the closest charged fermion to an object that can be observed in experiment. In particular, $\Theta(\Phi_5)=1$ if an event passes all selection cuts after an eventual photon recombination. 
The $\Theta(\tilde{\Phi}_{4,ij})$ only depend on an effective four-body kinematics if 
a photon recombination is applied as assumed in \citere{Dittmaier:1999mb}; 
in the case of non-collinear-safe observables (not discussed in this paper) this four-body kinematics has to be
changed to a related five-body kinematics upon revoking the collinear splitting
(see \citere{Dittmaier:2008md} for details).
To allow for an exact cancellation of IR divergences, in the soft and collinear limits the cut functions must behave as $\Theta(\Phi_{5})\to \Theta(\tilde{\Phi}_{4,ij})$.   

The singular part of the real-emission contributions is obtained by integrating the subtraction terms over the regions of phase space where the photon becomes soft or collinear to a charged fermion. This can most conveniently be achieved by decomposing the full five-body phase space into a four-body phase space related to the kinematics of the subtraction functions 
and the phase space of the photon, 
\beq
\int_5 \rd\Phi_5 =
\int_0^1 \rd x
\int_4 \rd\tilde{\Phi}_{4,ij}(x) 
\int_1 \rd\Phi_{\gamma,ij}(x)\,.
\eeq
In order to regularize the singular contributions, 
in addition to the infinitesimal mass parameter $\lambda$ for the photon 
we need to assign small masses $m_i$ to the fermions. 

The desired singular part can then be written in the form 
\beqar
\label{eq:int-counter}
\int_5
\rd\hat{\sigma}_{\bar qq}^\mr{real,sing}
&=&
-\frac{\alpha}{2\pi}
\sum_{i,j=1\atop i\neq j}^6
(-1)^{i+j}Q_i Q_j 
\non
\\
&&
\times
\int_0^1 \rd x
\left[
	\int_4 \rd\tilde{\Phi}_{4,ij}(x)\,
	\mc{G}_{ij}(\tilde{s}_{ij},x)
	F(x x_1 x_2)
	\sumav\left|\mc{M}_{\mr{Born}}(\tilde{\Phi}_{4,ij}(x))\right|^2
	\Theta(\tilde{\Phi}_{4,ij}(x))
\right.
\non\\
&&
\hs{8ex}
-
\left.
	\int_4 \rd\tilde{\Phi}_{4,ij}(1)\,
	\mc{G}_{ij}(\tilde{s}_{ij},x)
	F(x_1 x_2)
	\sumav\left|\mc{M}_{\mr{Born}}(\tilde{\Phi}_{4,ij}(1))\right|^2
	\Theta(\tilde{\Phi}_{4,ij}(1))
\right]
\non
\\
%
%
&&
{}-
\frac{\alpha}{2\pi}
\sum_{i,j=1\atop i\neq j}^6
(-1)^{i+j}Q_i Q_j 
F(x_1 x_2)
\int_4 \rd\Phi_{4}\,
G_{ij}(s_{ij})
\sumav\left|\mc{M}_{\mr{Born}}(\Phi_{4})\right|^2
\Theta(\Phi_{4})\,,
\eeqar
where the $\mc{G}_{ij}$ functions stem from the integral of the respective radiator functions over the phase space of the photon,
\beq
\mc{G}_{ij}(\tilde{s}_{ij},x)
=
8\pi^2 x
\int_1 \rd\Phi_{\gamma,ij}(x)\,
g_{ij}(q_i,q_j,k)\,.
\eeq
Momenta that are obtained by mapping the full five-body phase space onto $\tilde{\Phi}_{4,ij}$ are denoted as $\tilde{q}_i$, $\tilde{q}_j$, etc., such that $\tilde{s}_{ij}=2\tilde{q}_i\cdot\tilde{q}_j$.  
The endpoint contributions are contained in the last line of Eq.~(\ref{eq:int-counter}). The $G_{ij}$ functions are obtained by integrating the $\mc{G}_{ij}$ over $x$, 
\beq
G_{ij}(s_{ij}) = \int_0^1 \rd x \, \mc{G}_{ij}(s_{ij},x)
=\mc{L}(s_{ij},m_i^2) + C_{ij},
\eeq
with the function $\mc{L}$ and the constants $C_{ij}$ of \refeq{eq:l-functions} and \refeq{eq:Cij}, respectively.
Here, again we consistently combined the functions $\mc{G}_{ij,\pm}$ and $G_{ij,\pm}$ of
\citere{Dittmaier:1999mb} for polarized fermions
into $\mc{G}_{ij}=\mc{G}_{ij,+}+\mc{G}_{ij,-}$ and $G_{ij}=G_{ij,+}+G_{ij,-}$ for unpolarized situations.

In the case of final-state emitter--spectator pairs, $\mc{G}_{ij}$ is proportional to $\delta(1-x)$, 
and therefore only the last term of Eq.~(\ref{eq:int-counter}) survives. 
For a final-state emitter $i$ and an initial-state spectator $j$, the $\mc{G}_{ij}$ are 
actually independent of $\tilde{s}_{ij}$ and take the form  
\beq
\mc{G}_{ij}(\tilde{s}_{ij},x) = 
\frac{1}{1-x}
\left[
2\ln\left(
\frac{2-x}{1-x}
\right)
-\frac{3}{2}
\right], \qquad i=3,\dots,6, \quad j=1,2.
\eeq
For an initial-state emitter $i$ and a final-state spectator $j$ the distributions read
\beqar
\label{eq:qq-int-counter-if}
\mc{G}_{ij}(\tilde{s}_{ij},x) &=&
P_{ff}(x)
\left[
\ln\left(
\frac{\tilde{s}_{ij}}{m_i^2 x}
\right)
-1
\right]
-\frac{2}{1-x}
\ln(2-x)
+(1+x)\ln(1-x)+1-x\,,
\nn\\
&& i=1,2, \quad j=3,\dots,6,
\eeqar
with the fermion-to-fermion splitting function
\beq
P_{ff}(x) = \frac{1+x^2}{1-x}\,.
\eeq
Obviously, the $\mc{G}_{ij}$ distributions in this case explicitly depend on the fermion-mass parameter $m_i$ 
which is used to regularize the collinear singularity related to the splitting of the incoming fermion. 

When both emitter and spectator are initial-state particles, the integrated counterterms take the form 
\beq
\label{eq:qq-int-counter-ii}
\mc{G}_{ij}(s_{ij},x) = 
P_{ff}(x)
\left[
\ln\left(
\frac{s_{ij}}{m_i^2}
\right)
-1
\right] + 1-x,
\qquad i,j=1,2
\eeq

Eventually, the singularities appearing in the $\mc{G}_{ij}$
are taken care of by a {\em factorization term}. For the subprocess 
$\bar qq\to \mvev \gamma$ the term corresponding to a redefinition of the 
antiquark distribution function $f_{\bar q}$ is of the form 
\beqar
\label{eq:qq-fact}
\lefteqn{\int_0^1 \rd x_1
\int_0^1 \rd x_2\,
f_{\bar q}(x_1,\muf)  f_{q}(x_2,\muf) \, 
\int_0^1 \rd x \int_4
\rd\hat{\sigma}_{\bar qq}^{\mr{fact},\bar q} } &&
\nn\\
&=&
-\frac{\alpha}{2\pi}Q_q^2
\int_0^1 \rd x_1
\int_0^1 \rd x_2\,
F(x_1 x_2)
\int_{x_1}^1
\frac{\rd z}{z} \,
f_{\bar q}\left(\frac{x_1}{z},\muf\right)  f_{q}(x_2,\muf) 
\non\\
&&
{}\times
\left\{
	\ln\left(\frac{\muf^2}{m_q^2}\right)
	\left[P_{ff}(z)\right]_+
	-
	\left[
	P_{ff}(z)\left(
	2\ln(1-z)+1\right)
	\right]_+
	+C_{ff}^\mr{FS}(z) \right\}
	\non\\
	&& {}\times
\int_4\rd\Phi_4\,
\sumav\left|\mc{M}_{\mr{Born}}^{\bar qq\to \mvev}(x_1,x_2)\right|^2\, ,
\eeqar
where for clarity the arguments $(x_1,x_2)$ in the matrix element
indicate the momentum fractions of the incoming partons of the hard scattering process
relative to the proton momenta. 
The factorization term resulting from the redefinition of the quark PDF follows
from obvious substitutions.
The explicit form of the coefficient functions $C_{ff}^\mr{FS}$ depends on the 
{\it factorization scheme} used to define the PDFs (see \citeres{Spiesberger:1994dm,Diener:2005me} and references therein), 
where the currently available PDF sets with QED corrections, MRSTQED2004~\cite{Martin:2004dh} and
NNPDF2.3QED~\cite{Ball:2013hta}, are most appropriately used in the DIS scheme 
for the QED corrections.
In the DIS and $\overline{\mathrm{MS}}$ schemes the coefficient functions read
\beqar
C_{ff}^\mr{\overline{\mr{MS}}}(z) &=& 0\,,
\\
C_{ff}^\mr{DIS}(z) &=& \left[
P_{ff}(z)\left(
\ln\left(\frac{1-z}{z}\right)-\frac{3}{4}\right)+
\frac{9+5z}{4}
\right]_+\,.
\eeqar
Here, we are making use of the so-called ``plus prescription'' $[\ldots]_+$ 
which is defined via its action within an integral, 
\beq
\int_0^1 \rd z \, [g(z)]_+ f(z) = 
\int_0^1 \rd z \, g(z)\left[f(z)-f(1)\right]\,.
\eeq
An inspection of the expressions for the integrated counterterms in 
Eqs.~(\ref{eq:qq-int-counter-if}) and (\ref{eq:qq-int-counter-ii}) 
and the factorization contribution, Eq.~(\ref{eq:qq-fact}), reveals that the dependence on logarithms 
of the regulator mass completely drops out 
in the sum of singular and factorization contributions to the hadronic cross section.


\subsubsection{Subprocesses with incoming photons}
In subprocesses with an incoming photon, such as $\gamma q\to \mvev q$, singularities arise when the incoming photon 
splits into a collinear quark--antiquark pair, as illustrated on the l.h.s.\ of \reffi{fig:gamma-splitting} with $f=q$.
%
\bfi
\centerline{
\begin{picture}(260,110)(0,0)
\put(0,-10){
  \begin{picture}(100,120)(0,0)
  \Photon(20,90)(50,70){2}{4}
  \Line(50,70)(80,50)
  \Line(80,50)( 40, 20)
  \LongArrow( 52,18)( 67, 30)
  \LongArrow( 65,70)( 80, 80)
  \LongArrow( 20,80)( 35, 70)
  \Line(50,70)(80,90)
  \Vertex( 50,70){2.5}
  \GCirc(80,50){10}{1}
  \put( 10, 88){$\gamma$}
  \put( 87, 90){$f$}
  \put( 52, 50){$\bar f$}
  \put( 28, 18){$a$}
  \put( 65, 18){$q_a$}
  \put( 20, 63){$k$}
  \put( 76, 68){$q_f$}
  \end{picture} } 
\put(140,-10){
  \begin{picture}(100,120)(0,0)
  \Line(20,90)(50,70)
  \Photon(50,70)(80,50){2}{4}
  \Photon(80,50)( 40, 20){2}{5.5}
  \LongArrow( 52,18)( 67, 30)
  \LongArrow( 65,70)( 80, 80)
  \LongArrow( 20,80)( 35, 70)
  \LongArrow( 53.5,57.5)( 62.5, 52.5)
  \Line(50,70)(80,90)
  \Vertex( 50,70){2.5}
  \GCirc(80,50){10}{1}
  \put(  8, 88){$q$}
  \put( 87, 90){$q$}
  \put( 45, 57){$\gamma$}
  \put( 28, 18){$\gamma$}
  \put( 65, 18){$q_\gamma$}
  \put( 20, 63){$q_q$}
  \put( 78, 68){$q'_q$}
  \put( 53, 45){$k$}
  \end{picture} } 
\end{picture} }
\caption{Sketch of the splittings $\gamma\to f\bar f^\star$ (left)
and $f\to f\gamma^\star$ (right), both with an initial-state spectator ($a$ on the l.h.s. and $\ga$ on the r.h.s.).}
\label{fig:gamma-splitting}
\efi
This collinear singularity can be isolated via a ``dipole-subtraction inspired'' counterterm \cite{Dittmaier:2008md} 
constructed from the tree-level matrix elements for the underlying subprocess $\bar q q \to \mvev$. 
In principle, a collinear splitting of the incoming photon into a lepton pair $\ell\bar\ell$ can occur as well, 
c.f.\ l.h.s.\ of \reffi{fig:gamma-splitting} with $f=\ell$. 
However, as soon as we require the two final-state charged leptons to have non-vanishing transverse momenta, 
such configurations lie outside the relevant phase space. We therefore do not consider a collinear $\gamma\to\ell\bar\ell$ splitting in the following. 
An additional class of divergences arises when the incoming quark splits into a collinear photon--quark pair, as sketched on the r.h.s.\ of \reffi{fig:gamma-splitting}.  
%
%
Those singularities 
can be removed by counterterms constructed from the spin-correlated matrix elements for the underlying $\gamma\gamma \to \mvev$ subprocess~\cite{Dittmaier:2008md}. However, we opt for the alternative, simpler
procedure of multiplying the full real-emission matrix element for $\gamma q\to \mvev q$ with a so-called {\em effective collinear factor} that restores the correct quark-mass dependence of the
asymptotic behaviour 
in the phase-space region where the incoming and outgoing quarks become collinear. 

Let us first turn to the case of the incoming photon splitting into a quark--antiquark pair. 
Although in all configurations considered in this work the incoming and the outgoing quark of the real-emission process are of the same type, we label them differently to emphasize the splitting sequence. As illustrated in \reffi{fig:gamma-splitting} (l.h.s.), 
we denote the incoming spectator fermion by $a$, and the outgoing parton resulting from the splitting of the incoming photon by $f$, such that the entire subprocess reads $\gamma a \to \mvev\, f$. The collinear singularity related to a photon splitting of the form $\gamma\to f\bar f^\star$ is taken care of 
by a counterterm constructed from the tree-level matrix elements for the underlying subprocess $\bar f a \to \mvev$
\cite{Dittmaier:2008md},
\beqar 
\sumav\left|\mc{M}_{\mr{sub},\gamma a}\right|^2
=
N_f^{\mathrm{c}}
\,Q_f^2 \, e^2\,
h^{\gamma f,a}(k,q_f,q_a)
\sumav\left|\mc{M}_{\mr{Born}}^{\bar f a \to \mvev}(\tilde{\Phi}_{4,\gamma a})\right|^2\,,
\label{eq:Msubga}
\eeqar
where $h^{\gamma f,a}$ is the the radiator function
\beq
h^{\gamma f,a}(k,q_f,q_a)
= 
\frac{2}{x_{f,\gamma a}s_{f\gamma}}
P_{f\gamma}(x_{f,\gamma a}),
\eeq
with 
\beq
x_{f,\gamma a} = \frac{s_{a\ga}-s_{f\gamma}-s_{af}}{s_{a\ga}}
\eeq
and the photon-to-fermion splitting function
\beq
P_{f\gamma}(x) = (1-x)^2+x^2\,. 
\eeq
The colour factor $N_f^{\mathrm{c}}$ of the fermion $f$, with $N_q^{\mathrm{c}}=3$ and $N_\ell^{\mathrm{c}}=1$,
appears on the r.h.s.\ of Eq.~\refeq{eq:Msubga} because of the colour average in $\sumav$. 
Note that we again simplified the construction of the counterterm upon specializing the 
formulas for $h^{\gamma f,a}_\pm$ describing polarized photons~\cite{Dittmaier:2008md} 
to the unpolarized case, where only $h^{\gamma f,a}=h^{\gamma f,a}_++h^{\gamma f,a}_-$
is relevant.
The finite part of the real-emission contribution for a photon-induced subprocess is then of the form
\beq
\label{eq:gaq-fin}
\int_5
\rd\hat{\sigma}_{\gamma q}^\mr{real,fin} 
=
F(x_1 x_2)
\int_5 \rd\Phi_5 \left[
\sumav\left|\mc{M}_{\mr{real}}^{\gamma q \to\mvev q}\right|^2\Theta(\Phi_5)
-
\sumav\left|\mc{M}_{\mr{sub},\gamma a}\right|^2
\Theta(\tilde{\Phi}_{4,\gamma a})
\right]\,.
\label{eq:sigmagqrealfin}
\eeq
Analogous to the case of subprocesses with an incoming quark--antiquark pair the divergent part of the real-emission contribution for photon-induced processes is obtained by integrating over the singular regions of phase space. The respective contribution to the four-body cross section is given by 
\beqar
\int_5 \rd\hat{\sigma}_{\gamma q}^\mr{real,sing} 
&=&
N_f^{\mathrm{c}}
\frac{Q_f^2\alpha}{2\pi}
\int_0^1
\rd x\,
F(x x_1 x_2)
\int_4 \rd\tilde{\Phi}_{4,\gamma a}\,
\mc{H}^{\gamma f,a}(s_{\gamma a},x) \,
\nn\\ && {} \times
\sumav\left|\mc{M}_{\mr{Born}}^{\bar f a \to \mvev}(\tilde{\Phi}_{4,\gamma a})\right|^2
\Theta(\tilde{\Phi}_{4,\gamma a})
\label{eq:sigmagqrealsing}
\eeqar
with 
\beq
\mc{H}^{\gamma f,a}(s_{\gamma a},x)
=
P_{f\gamma}(x)
\ln\left[
\frac{s_{\gamma a}(1-x)^2}{m_f^2}
\right]
+2x(1-x).
\eeq
For each quark flavour 
$f=q$, the contributions of 
\refeq{eq:sigmagqrealfin} and \refeq{eq:sigmagqrealsing} are to be convoluted
with the respective PDFs $f_{\ga}(x_1,\muf)$ and $f_{q}(x_2,\muf)$. 
The factorization term that cancels the mass singularity in this expression is of the form~\cite{Dittmaier:2009cr}
\beqar
\lefteqn{\int_0^1 \rd x_1
\int_0^1 \rd x_2\,
f_{\ga}(x_1,\muf)  f_{q}(x_2,\muf) \, 
\int_0^1\rd x\,\int_4\rd\hat{\sigma}_{\gamma q}^\mr{fact}
} &&
\nn\\
&=&
-N_q^{\mathrm{c}} \frac{\alpha}{2\pi}Q_q^2
\int_0^1 \rd x_1
\int_0^1 \rd x_2\,
F(x_1 x_2)
\int_{x_1}^1
\frac{\rd z}{z} \,
f_{\ga}\left(\frac{x_1}{z},\muf\right)  f_{q}(x_2,\muf) 
\non\\
&&
{}\times
\Biggl\{
	\ln\left(\frac{\muf^2}{m_q^2}\right)
	P_{f\gamma}(z)
	+C_{f\gamma}^\mr{FS}(z)
\Biggr\}
\int_4\rd\Phi_4\,
\sumav
\left|\mc{M}_{\mr{Born}}^{\bar q q\to \mvev}(x_1,x_2)\right|^2\,, 
\eeqar
with 
\beqar
C_{f\gamma}^\mr{\overline{\mr{MS}}}(z) &=& 0\,,
\\
C_{f\gamma}^\mr{DIS}(z) &=& 
P_{f\gamma}(z) \ln\left(\frac{1-z}{z}\right)-8z^2+8z-1.
\eeqar

As mentioned above, singularities of the $\gamma q \to \mvev q$ amplitudes that are related to the collinear splitting 
of the incoming quark into a quark--photon pair are most conveniently being tackled by an effective collinear factor 
that multiplies the entire $\gamma q\to \mvev q$ cross section of Eq.~(\ref{eq:gaq-fin}), including the counterterms 
taking care of collinear  $\gamma\to f\bar f^\star$ splittings. 
This collinear factor has to be constructed in such a way that it restores the correct leading quark-mass dependence
of the $\gamma q \to \mvev q$ amplitude squared in the limit of forward quark scattering, but reduces to a factor~$1$ 
up to (negligible) mass-suppressed terms outside the collinear region.\footnote{The application of an effective collinear 
factor to photonic bremsstrahlung corrections is, e.g., 
described in \citere{Dittmaier:2001ay}.}
This factor can
be read off from Eq.~(5.6) of \citere{Dittmaier:2008md}, where the asymptotic factorization formula for the
splitting is given for a non-zero quark mass. The effective collinear factor is just the ratio
of the massive version of this formula to its massless version and explicitly reads
\beq
f(m_q,x,E_q,\theta) 
=
\frac{
\sin^2\frac{\theta}{2}}
{
\left[
\sin^2\frac{\theta}{2}+
\frac{m_q^2 x^2}{4 E_q^2(1-x)^2}
\right]^2
}
\left\{
\sin^2\frac{\theta}{2}+
\frac{m_q^2 x^4}{4 E_q^2(1-x)^2(1+(1-x)^2)}
\right\}\,,
\eeq
where $\theta$ is the angle between the incoming and the outgoing quark of mass $m_q$ defined in the partonic centre-of-mass frame, 
$E_q$ is the energy of the incoming quark in the same frame, and
$x=k^0/E_q$ is the fraction of the incoming quark's momentum that is carried by the emitted photon. 
The mass $m_q$ is supposed to be much smaller than $E_q$ and plays the role of a regulator for the collinear singularity. 
Multiplying the entire real-emission cross section for a massless quark $q$ with this factor renders the modified cross
section integrable over the collinear $q\to q\gamma^\star$ configurations, and the integral over the collinear region
reproduces the correct mass dependence up to mass-suppressed terms of order $\ord(m_q/E_q)$.

The dependence of the cross section on the regulator mass $m_q$
eventually drops out when an appropriate factorization term is added. 
As the collinear singularity related to the $q\to q\gamma^\star$ splitting can be assigned to the associated quark PDF, 
the appropriate factorization term for the $\gamma q \to \mvev q$ subprocess requires the partonic scattering amplitudes 
of the underlying $\gamma \gamma\to \mvev$ process~\cite{Dittmaier:2009cr},
\beqar
\lefteqn{\int_0^1 \rd x_1
\int_0^1 \rd x_2\,
f_{\gamma}(x_1,\muf)  f_{q}(x_2,\muf) \, 
\int_0^1 \rd x
\int_4 \rd\hat{\sigma}_{\gamma q}^\mr{fact}
} &&
\nn\\
&=&
-\frac{\alpha}{2\pi}Q_q^2
\int_0^1 \rd x_1
\int_0^1 \rd x_2\,
F(x_1 x_2)
\int_{x_2}^1
\frac{\rd z}{z}
f_{\gamma}\left(x_1,\muf\right)  f_{q}\left(\frac{x_2}{z},\muf\right) 
\non\\
&&
{}\times
\left\{
	\ln\left(\frac{\muf^2}{m_q^2}\right)
	P_{\gamma f}(z)
	-P_{\gamma f}(z)\left( 2\ln z +1\right)
	+C_{\gamma f}^\mr{FS}(z)
\right\}
\non\\
&&
{}\times
\int_4\rd\Phi_4\,
\sumav
\left|\mc{M}_{\mr{Born}}^{\gamma\gamma\to \mvev}(x_1,x_2)\right|^2,
\hspace*{2em}
\eeqar
with 
\beqar
C_{\gamma f}^\mr{\overline{\mr{MS}}}(z) &=& 0\,,
\\
C_{\gamma f}^\mr{DIS}(z) &=& -C_{ff}^\mr{DIS}(z)
=
- 
\left[
P_{ff}(z)\left(
\ln\left(\frac{1-z}{z}\right)-\frac{3}{4}\right)+
\frac{9+5z}{4}
\right]_+\,.
\eeqar 
%
%

%

%
\subsection{Improved Born approximation}
The double-pole approximation relies on an expansion of the entire four-fermion production cross section around the resonance region of the two weak-boson propagators in $\Pp\Pp\to \PWp\PWm\to 4$~fermions. While this approach provides an 
excellent description of the production cross section of two resonant weak bosons, it is not applicable close to and 
below the resonance region. 
Therefore for partonic centre-of-mass energies $\sqrt{\hat s}<2\MW+5$~GeV,
instead of the NLO EW corrected cross section,
we employ a so-called {\em improved Born approximation} (IBA) that captures a major part of the full radiative corrections to $\Pp\Pp\to \mvev$ without requiring an on-shell projection procedure.

As worked out in detail in \citeres{Dittmaier:1991np,Denner:2001zp} for the related case of four-fermion production at lepton colliders, universal radiative corrections stemming from running couplings or Coulomb singularities can be accounted for by suitable correction factors applied to parts of the Born cross section. Generally, the values of the electroweak input parameters used in a calculation depend on the input-parameter scheme. In the $G_\mu$ scheme, 
which is used in our quark--antiquark induced LO cross section,
the electromagnetic coupling $e$ is computed from the Fermi constant, $G_\mu$, via tree-level electroweak relations. 
The weak mixing angle, $\theta_\mr{W}$, is determined by the gauge-boson masses $\MW$ and $\MZ$ according to
\beq
\cos\theta_\mr{W}\equiv\cw=\sqrt{1-\sw^2}=\MW/\MZ. 
\label{eq:cw}
\eeq
In the IBA, the replacements
\beq
\frac{e^2}{\sw^2}
\to
4\sqrt{2} G_\mu \MW^2\,,
\quad
e^2
\to 
4\pi\alpha(\hat s)
\eeq
are made in the tree-level matrix elements, implying that weak-isospin exchange involves the coupling $G_\mu \MW^2$, and pure photon exchange the coupling $\alpha(Q^2)$ at a scale $Q^2$ set by the kinematics of the reaction. The running of the electromagnetic coupling is determined by the light charged fermions, so that
\beq
\alpha(Q^2) = 
\frac{\alpha(\MZ^2)}{
1-\frac{\alpha(\MZ^2)}{3\pi}
\ln\left(\frac{Q^2}{\MZ^2}\right)
\sum_{f\neq \Pt}  
N_f^{\mathrm{c}} Q_f^2}
\,,
\eeq
with the colour factor $N_f^{\mathrm{c}}=1$ or $3$ if the fermion $f$ is a lepton or quark, respectively.
The measured value of $\alpha(\MZ^2)$ is then taken as numerical input.

In addition to this prescription for the couplings, we include a universal correction factor $\delta_\mr{Coul}$ to account for the contribution of the Coulomb singularity stemming from photon exchange near the $\PW$-pair production threshold. The complete partonic IBA cross section  is then of the form 
\beq
\rd\hat{\sigma}_{\bar qq}^\mr{IBA} 
=
F(x_1 x_2)\,
\rd\Phi_4\,
\sumav\,
\left|
\mc{M}_\mr{IBA}^{\bar qq\to \mvev}
\right|^2
\left[
1+\delta_\mr{Coul}(\hat{s},k^2_+,k^2_-)
\right]
g(\bar\beta)\,.
\eeq
The Coulomb singularity is contained in the factor~\cite{Denner:1997ia,Fadin:1993kg}
\beqar
\delta_\mr{Coul}(\hat{s},k^2_+,k^2_-)
&=&
\frac{\alpha(0)}{\bar\beta}
\mr{Im}
\left\{
\ln\left(
\frac{
\beta-\bar\beta+\Delta_M}{
\beta+\bar\beta+\Delta_M
}
\right)
\right\}
\,,
\non\\
\bar\beta &=&
\frac{\sqrt{
\hat{s}^2+k^4_++k^4_--2\hat{s}k^2_+-2\hat{s}k^2_--2k_+^2k_-^2
}
}{\hat{s}
}\,,
\non\\
\beta &=&
\sqrt{1-\frac{4(\MW^2-\ri\MW\Gamma_\PW)}{\hat{s}}}\,,
\quad
\Delta_M =
\frac{
\left|
k_+^2-k_-^2
\right|
}{\hat{s}}
\eeqar 
with the fine-structure constant $\alpha(0)$ and a damping factor
\beq
g(\bar\beta) =
\left(
1-\bar\beta^2\right)^2\,,
\eeq
that ensures that $\delta_\mr{Coul}$ only has an impact in the threshold region.

%
\subsection{Default implementation and independent checks}
For our default implementation of the 
tree-level and real-emission contributions, discussed in \refses{ssec:born} and \ref{ssec:reals},
inhouse tree-level matrix elements expressed in terms of Weyl-van-der-Waerden spinor products are used.
The loop amplitudes are obtained with inhouse {\tt Mathematica} 
routines in complete analogy to the formulas used in {\tt RacoonWW}~\cite{Denner:2002cg}.
The numerical integration is performed by an adapted version of the multi-channel phase-space generator in the Monte Carlo
program {\tt Coffer$\ga\ga$}~\cite{Bredenstein:2005zk}, fully based on {\tt Fortran} code. 

For all parts of the calculation we worked out two independent implementations the results of which are in very good agreement,
both for matrix elements (squared) at
individual phase-space points and for cross sections after phase-space integration.  Tree-level matrix elements  
were compared to amplitudes generated with the {\tt MadGraph~4} package~\cite{Stelzer:1994ta,Alwall:2007st}.

The alternative implementation of the calculation which was used for comparing the
virtual amplitudes of \refse{ssec:virtuals} and phase-space integrated results
uses the {\tt FormCalc}~package~\cite{Hahn:1998yk} in order to calculate
analytic expressions for the tree-level and one-loop amplitudes of the
$2\rightarrow 2$ and $1\rightarrow 2$ processes (including the counterterms).
Inhouse {\tt Mathematica} code was used to transform these amplitudes into {\tt
Fortran} code which links to the runtime library of the matrix element generator
{\tt O'Mega}~\cite{Moretti:2001zz} and which assembles them into DPA amplitudes
as discussed in \refse{ssec:born} -- \ref{ssec:virtuals}. For phase-space
integration, a customized version of the {\tt Whizard}
package~\cite{Kilian:2007gr} was used which we augmented with an implementation
of the dipole subtraction formalism as discussed in \refse{ssec:reals}. The
tree-level amplitudes required for the real-emission and subtraction
contributions were generated by {\tt O'Mega}. 

%
\section{Phenomenological results}
\label{sec:pheno}
\subsection{Description of the setup}
\label{ssec:setup}
For our numerical studies we consider proton--proton collisions at the LHC at centre-of-mass energies of $\sqrt{\spp} = 14\TeV$
(``LHC14'') and 8~TeV (``LHC8''), respectively. We are using the SM input parameters
\beq
\begin{array}{rlrlrlrl}
G_\mu &= 1.1663787\times10^{-5}\GeV^{-2}, &
\\
\alpha(0) &=1/137.035999074, &
\alpha(\MZ) &= 1/128.93, &&
\\
\MW^\OS &=80.385\GeV, &
\Gamma_\PW^\OS & =2.085\GeV, 
\\
\MZ^\OS &=91.1876\GeV, &
\Gamma_\PZ^\OS &=2.4952\GeV,
\\
\MH &=125.9\GeV, &
\\
\Mt&=173.07\GeV, & \Mb&=4.78\GeV,
\end{array}
\eeq
essentially following \citere{Beringer:1900zz}

The DPA prescription of the W~resonances, which is used for the virtual NLO EW corrections, and
the complex-mass scheme for the W and Z~resonances, which is used for the LO predictions and for the
real corrections, both employ the pole definition of the gauge-boson masses with fixed decay widths
in the corresponding propagators.
At LEP and the Tevatron, however, running-width prescriptions were used to fit the W~and Z~resonances,
i.e.\ the respective masses correspond to the ``on-shell'' (OS) values given above.
We, thus, convert these OS
values $M_V^{\OS}$ and $\Ga_V^{\OS}$ ($V=\PW,\PZ$), resulting
from LEP and Tevatron, to the ``pole values'' denoted by $M_V$ and $\Ga_V$
according to \citere{Bardin:1988xt},
\beq\label{eq:m_ga_pole}
M_V = M_V^{\OS}/
\sqrt{1+(\Ga_V^{\OS}/M_V^{\OS})^2},
\qquad
\Ga_V = \Ga_V^{\OS}/
\sqrt{1+(\Ga_V^{\OS}/M_V^{\OS})^2},
\eeq
leading to
\beqar
\begin{array}[b]{r@{\,}l@{\qquad}r@{\,}l}
\MW &= 80.357\ldots\GeV, & \GW &= 2.0842\ldots\GeV, \\
\MZ &= 91.153\ldots\GeV,& \GZ &= 2.4943\ldots\GeV.
\label{eq:m_ga_pole_num}
\end{array}
\eeqar
These mass and width parameters are used in the numerics
discussed below, although the difference between using 
$M_V$ or $M_V^{\OS}$ would be hardly visible.
The weak mixing angle $\theta_W$ is computed from the masses of the weak gauge bosons via
Eq.~\refeq{eq:cw} in the (finite) virtual corrections; at LO and for the finite real corrections
its complex version $\cw^2=(\MW^2-\ri\MW\GW)/(\MZ^2-\ri\MZ\GZ)$ is used.
As mentioned before, we determine electroweak parameters in the $G_\mu$ scheme with the effective coupling
\beq
\alpha_{G_\mu} = \frac{\sqrt{2}\,G_\mu \MW^2 \sw^2}{\pi}
\eeq
in the LO matrix elements, with $\sw$ as in Eq.~\refeq{eq:cw}. By this input procedure, all higher-order contributions associated with the running of the 
electromagnetic coupling up to the electroweak scale and the leading universal two-loop top-mass dependent corrections 
to the $\rho$-parameter are absorbed in the LO cross section. 
In this scheme, no significant dependence of the predictions on the light-fermion masses results, although
we keep the full fermion-mass dependence in internal closed fermion loops. It is, therefore, sufficient to
give the values of the largest fermion masses, $\Mt$ and $\Mb$, above, while the other fermion masses are
numerically irrelevant. Recall that in this paper we consider only observables for ``dressed'' charged leptons,
so that no lepton-mass dependence from collinear final-state radiation remains either. Whenever quark or
lepton masses are used as small regulator masses, no numerically relevant impact of their values remains in our results.

We use two different values for the electromagnetic coupling $\alpha$ in the {\it relative}
NLO EW corrections, which are of $\ord(\alpha)$.
Specifically, we set $\alpha=\alpha_{G_\mu}$ in the finite virtual corrections, which do not involve any
leading contribution from photon radiation, but comprise all leading EW high-energy logarithms that are potentially
large at high scales. In the remaining EW corrections we take  $\alpha=\alpha(0)$, since those are entirely furnished
by photonic effects, for which $\alpha(0)$ is the appropriate effective coupling.
We expect that this setting is most stable against universal higher-order EW corrections connected with 
coupling renormalization~\cite{Denner:1991kt}.

For the parton distribution functions of the proton, we use the NNPDF2.3QED~\cite{Ball:2013hta} set which
supports QED effects at NLO, including a distribution function for the photons inside the proton. 
Our default choice for the factorization scale is the mass of the $\PW$~boson, 
\beq
\muf = \MW\,.
\eeq
Since the presented EW corrections show a very weak scale dependence, as illustrated below, we refrain from introducing specifically
adjusted dynamical scale choices as a default. 

In order to obtain IR-safe observables for the process $\Pp\Pp\to \mvev +X$ we employ a recombination procedure 
that combines final-state leptons and nearly collinear photons to pseudo-particles, in line with the
concept of ``dressed leptons'' used by ATLAS (see e.g.\ \citere{Aad:2011gj}).
In detail, we first determine the separation $R_{ij}$ of two particles $i$ and $j$ in the rapidity--azimuthal-angle plane, 
\beq
R_{ij} = 
\sqrt{
(y_i-y_j)^2+(\Delta\phi_{ij})^2
}\,,
\eeq
where $y_k$ denotes the rapidity and $\Delta\phi_{ij}$ the azimuthal angle between particles $i$ and $j$. 
In this process, we only consider photons with rapidities 
\beq
|y_\gamma| < 5\,.
\eeq
Photons inside this rapidity region are recombined with the closest charged lepton~$\ell$ whenever
\beq
R_{\gamma \ell}<0.1\,.
\eeq
For such configurations, we form pseudo-leptons by simply adding the momenta of the photon and the respective lepton. 
The momenta of the other particles in the event remain unaffected.
\subsubsection*{(i)~Default set of cuts}
Unless stated otherwise, for an acceptable event 
we require the presence of two oppositely charged leptons with a transverse momentum 
\beq
\label{eq:ptl-cut}
p_{\rT,\ell} > 20\GeV\,,
\eeq
in the central-rapidity region of the detector,
\beq
\label{eq:yl-cut}
|y_\ell| < 2.5\,.
\eeq

In subprocesses with a final-state quark or antiquark, this parton $i$ is considered as a jet
if its transverse momentum $p_{\rT,i}$ is larger than 
\beq
\label{eq:jet-pt}
p_\mr{T,jet}^\mr{cut} = 25\GeV,
\eeq
and its rapditiy lies within the range accessible to the detector,
\beq
|y_i|<5\,.
\eeq
If the parton does not meet these two requirements, the event is considered as part of the zero-jet cross section. 

In addition, we require that a hard parton~$i$ with $p_{\rT,i}>p_\mr{T,jet}^\mr{cut}$ be well-separated from the 
charged leptons and discard events with a hard final-state parton, unless
\beq
R_{i\ell}>0.4\,.
\eeq 
In order to avoid overwhelmingly large QCD corrections 
with a very asymmetric configuration of 
the lepton system, we impose a veto rejecting any event with a hard jet of
\beq
\label{eq:jet-veto}
p_\mr{T,jet}>100\GeV\,.
\eeq
We anticipate that this restriction also suppresses the impact of 
quark--photon initiated subprocesses, which are kinematically related to
quark--gluon scattering, to the level of a few percent.

\subsubsection*{(ii)~ATLAS cuts}
It is interesting to consider distributions in a more realistic setup, with selection cuts inspired by a recent analysis of the ATLAS collaboration for $\Pp\Pp\to \PWp\PWm$~\cite{ATLAS:2012mec}. To this end, in addition to the lepton cuts and recombination criteria of our default setup we impose 
transverse-momentum cuts on the hardest charged lepton and the neutrino system, 
\beqar
p_{\rT,\ell}^\mr{leading} > 25\GeV,
\qquad
E_\rT^\mr{miss} = |\vec{p}_\rT^\mr{\;miss}| > 25\GeV,
\eeqar
and require the two charged leptons to be  
well separated from each other,
\beqar
R_{\Pe\mu}>0.1\,,
\qquad
M_{\Pe\mu} > 10\GeV.
\eeqar
Furthermore, we veto all real-emission events with a jet harder than
\beq
\label{eq:atlas-jveto}
p_{\rT,\mr{jet}}>25\GeV.
\eeq
%

%
\subsection{Numerical results}
The quark-initiated contributions to the LO cross section for $\Pp\Pp\to \mvev$ at the LHC both with $\sqrt{\spp}=14\TeV$ and with $\sqrt{\spp}=8\TeV$ are given in \refta{tab:xsecs} for our default setup and for our ATLAS setup in the latter case.
In the following, we 
employ $\sigma_{\bar q q}^\mr{LO}$ as our benchmark cross section and consider tree-level contributions with 
$\bar\Pb\Pb$ or $\gamma\gamma$ initial states as well as electroweak NLO contributions as corrections.
\begin{table}
\begin{center}
\begin{tabular}{|c|c|cccc|}
\hline
&$\sigma_{\bar q q}^\mr{LO}$~[fb] &  $\delta_{\bar q q}$~[$\%$]& $\delta_{q \gamma}$~[$\%$] & $\delta_{\gamma\gamma}$~[$\%$] & $\delta_{\bar\Pb\Pb}$~[$\%$]
\\
\hline
\hline
LHC14     &412.5(1) & $-2.70(2)$ & $0.566(5)$   & $0.7215(4)$  & $1.685(1)$ 
\\
\hline
LHC8      &236.83(5) & $-2.76(1)$ & $0.470(3)$   & $0.8473(3)$  & $0.8943(3)$
\\
\hline
ATLAS cuts&163.84(4) & $-2.96(1)$ & $-0.264(5)$  & $1.0221(5)$  & $0.9519(4)$ 
\\
\hline
\end{tabular}
\caption{
\label{tab:xsecs}
Cross-section contributions to $\Pp\Pp\to \mvev$ at the LHC running at $14\TeV$ (first line) and $8\TeV$ (second line), respectively, with the default settings of Sec.~\ref{ssec:setup}. The third line shows the corresponding results for a collider energy of $8\TeV$ with the ATLAS setup defined in Sec.~\ref{ssec:setup}. The numbers in brackets represent the numerical error on the last given digit.  
}
\end{center}
\end{table}
%
%
The cross section at LO accuracy is dominated by contributions initiated by light quarks, $\sigma_{\bar q q}^\mr{LO}$. Subprocesses with bottom quarks in the initial state yield an additional contribution of less than 2\% for the three setups under investigation. 
Only about 1\% of the full tree-level cross section stems 
from the photon-initiated contributions, $\sigma_{\gamma\gamma}^\mr{LO}$.
This result rectifies a posteriori the neglect of order $\mc{O}(\alpha)$  corrections to subprocesses of the type $\bar\Pb\Pb \to\mvev$ 
and $\gamma\gamma\to\mvev$.

The NLO EW 
cross section of Eq.~(\ref{eq:sigma-nlo}) contains virtual EW and photonic 
real-emission corrections to $\bar qq\to\mvev$, and the additional classes of (anti)quark--photon, photon--photon, and $\bar\Pb\Pb$
initiated subprocesses%
\footnote{For notational convenience, in the following we refer to the sum of 
$\ga q/q\gamma\to\mvev q$ and $\ga \bar q/\bar q \gamma\to\mvev\bar q$ channels
globally as ``quark--photon induced processes'', implicitly including the antiquark--photon contributions.}, 
$\sigma_{q\gamma}$, $\sigma_{\gamma\gamma}$, and $\sigma_{\bar\Pb\Pb}$, respectively.
Apparently, the sum of all considered corrections is very small as the small negative EW corrections to the quark-initiated processes are widely
compensated by positive corrections of the separately considered tree-level contributions. However, the EW corrections significantly distort 
distributions, as they are not uniformly distributed in phase space, but tend to increase at scales above the weak-boson mass. 

This feature of the EW corrections is clearly illustrated by the transverse momentum distribution of the electron in \reffi{fig:pte}~(left). 
%
\bfi
\includegraphics[angle=0,scale=0.7,bb=60 330 370 770]{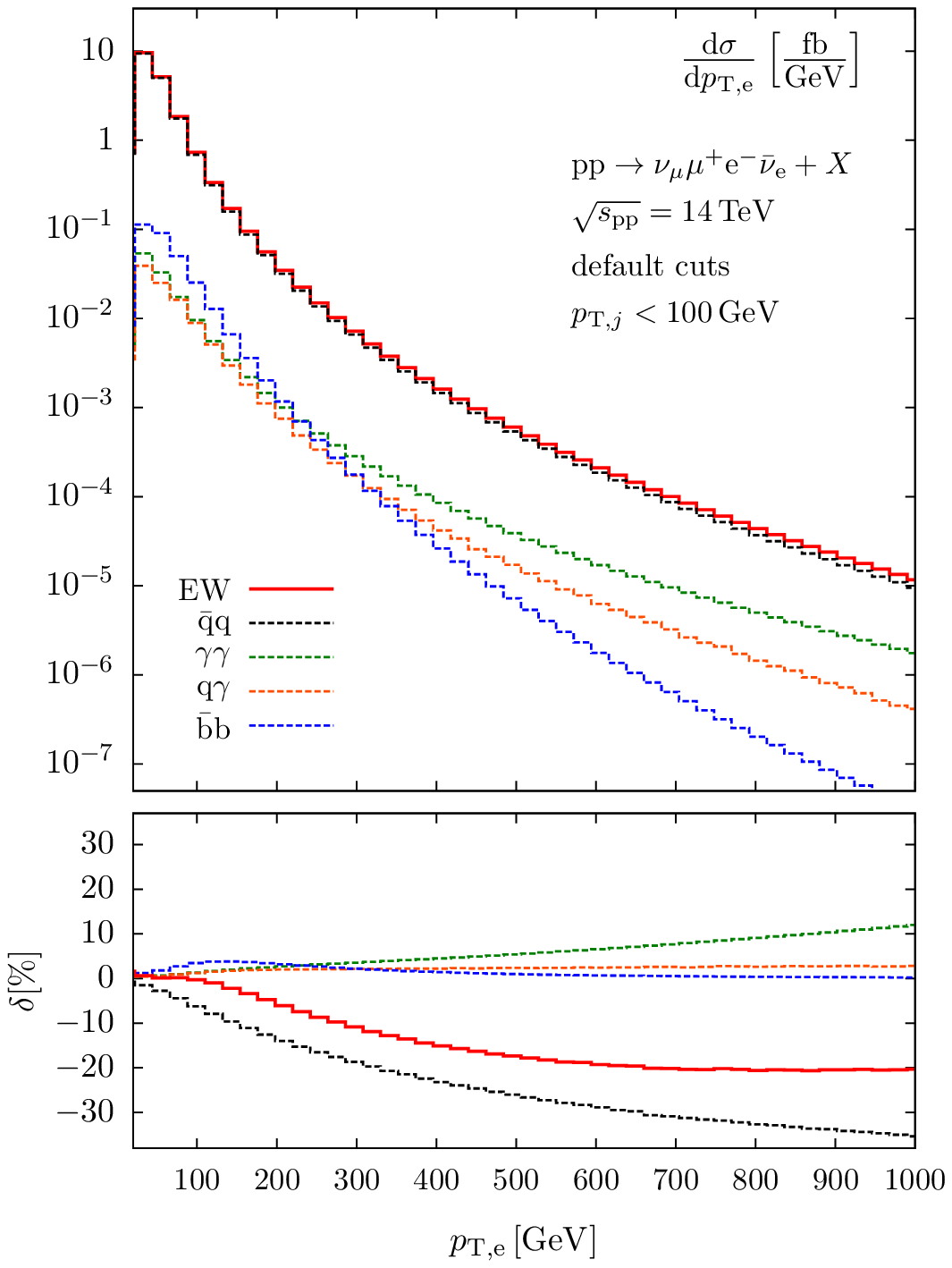}
\includegraphics[angle=0,scale=0.7,bb=60 330 370 770]{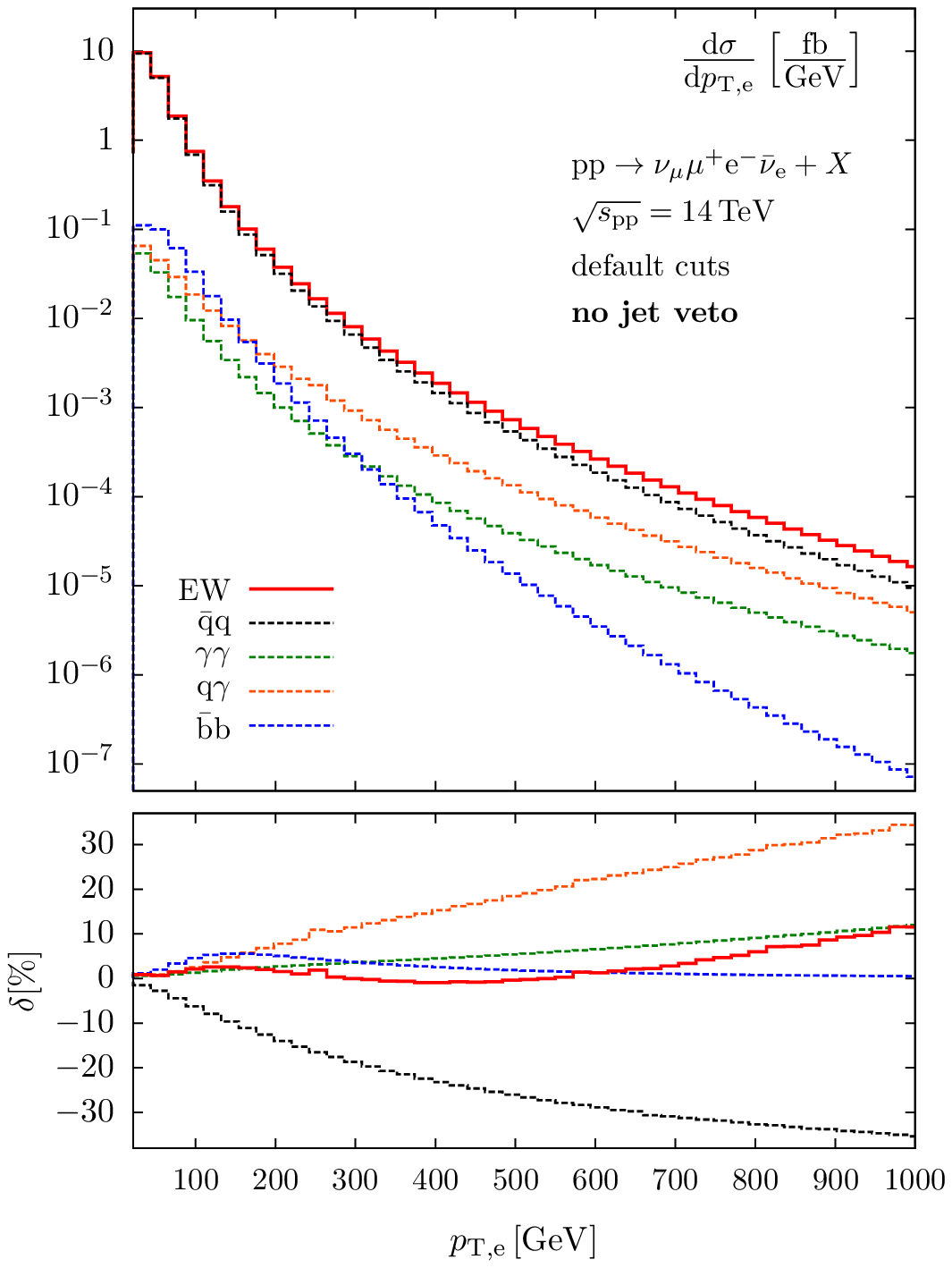}
\caption{Transverse-momentum distribution of the electron in $\Pp\Pp\to\mvev$ at NLO~EW accuracy (upper panels) for our default setup at the LHC14 with a jet veto of $100\GeV$ (left) and without a jet veto (right),  
together with the relative impact of individual contributions in each case (lower panels). 
\label{fig:pte}
}
\efi
%
The relative corrections of type $i$ normalized to the $\bar q q$-induced  LO
result for a specific distribution $\rd\sigma/\rd\mc{O}$, 
\beq
\delta_i(\mc{O}) = 
\frac{
	\frac{\rd\sigma^i}{\rd\mc{O}}-
	\frac{\rd\sigma_{\bar q q}^\mr{LO}}{\rd\mc{O}}
	}		
	{\frac{\rd\sigma_{\bar q q}^\mr{LO}}{\rd\mc{O}}
	}\,,	
\eeq
are displayed in the lower panels of the figure. Obviously, the NLO EW contributions induced by $\bar qq$ initial states become large and negative with increasing $p_{\rT,\Pe}$. Even though this effect is slightly balanced by an increase of $\rd\sigma_{q\gamma}/\rd p_{\rT,\Pe}$ with  $p_{\rT,\Pe}$, the full NLO~EW corrections considerably decrease the tail of the transverse-momentum distribution, amounting to more than $-10\%$
for $p_{\rT,\Pe}=300\GeV$. For even higher transverse momenta the $\gamma\gamma$-induced tree-level processes gain in relative importance, amounting to as much as 10\% when $p_{\rT,\Pe}$ approaches the TeV regime. 
The $\bar\Pb\Pb$ contributions, on the other hand, are relevant only at low $p_{\rT,\Pe}$, becoming completely insignificant above
$200\GeV$. 

We note that a much stronger increase of the quark--photon induced contributions at large $p_{\rT,\Pe}$ is found in the absence of the jet veto of Eq.~(\ref{eq:jet-veto}). This feature is illustrated by \reffi{fig:pte}~(right) 
that shows  $\rd\sigma_{q\gamma}/\rd p_{\rT,\Pe}$ for the same setup as \reffi{fig:pte}~(left), apart from the jet veto of Eq.~(\ref{eq:jet-veto}). Normalization and shape of all contributions that do not contain a QCD 
parton in the final state, and therefore cannot give rise to a jet, are identical to the case where a jet veto is imposed. However, a significant increase in the relative size of the quark--photon contributions can be observed in the tail of the transverse-momentum distribution, giving rise to a relative contribution $\delta_{q\gamma}$ of about
30\% for $p_{\rT,\Pe}=0.9\TeV$. 
This behaviour 
can be traced back to a mechanism referred to as ``giant $K$~factor'' in the literature~\cite{Rubin:2010xp}. Processes that involve weak bosons often exhibit QCD 
radiative corrections that dramatically grow at scales far above the boson mass and can approach several hundreds of percent, for instance, in the tails of transverse-momentum distributions. This effect is due to topologies that first occur at NLO and involve the emission of gauge bosons that are quasi-soft at the considered scales, thus giving rise to double-logarithmic corrections that grow at scales above the gauge-boson mass. Such a behaviour  
has been observed in related processes such as 
$\PW$+jet~\cite{Denner:2009gj} and $\PZ$+jet production~\cite{Denner:2011vu}, and has also been reported in \citere{Bierweiler:2012kw,Baglio:2013toa} for on-shell $\PW$-pair production at the LHC. In order to avoid these large contributions, in the following we 
therefore always impose the jet veto of Eq.~(\ref{eq:jet-veto}) and disregard events with a very hard jet. The jet veto reduces the contribution of the photon--quark induced channels to the integrated NLO cross section for our default setup from $1.1\%$ to $0.6\%$, the value given in \refta{tab:xsecs} for LHC14. 

%
\bfi
\includegraphics[angle=0,scale=0.7,bb=60 330 370 770]{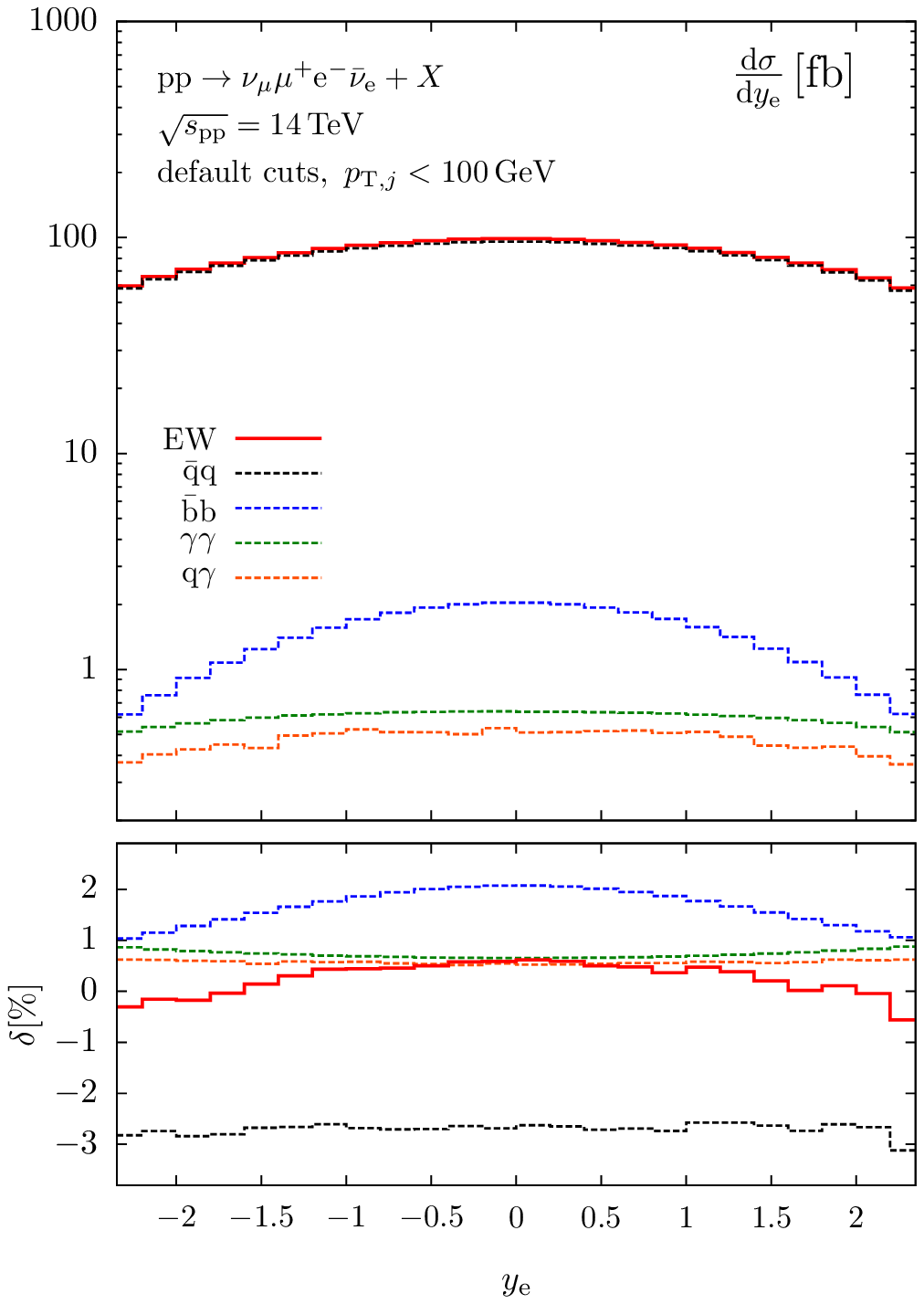}
\includegraphics[angle=0,scale=0.7,bb=60 330 370 770]{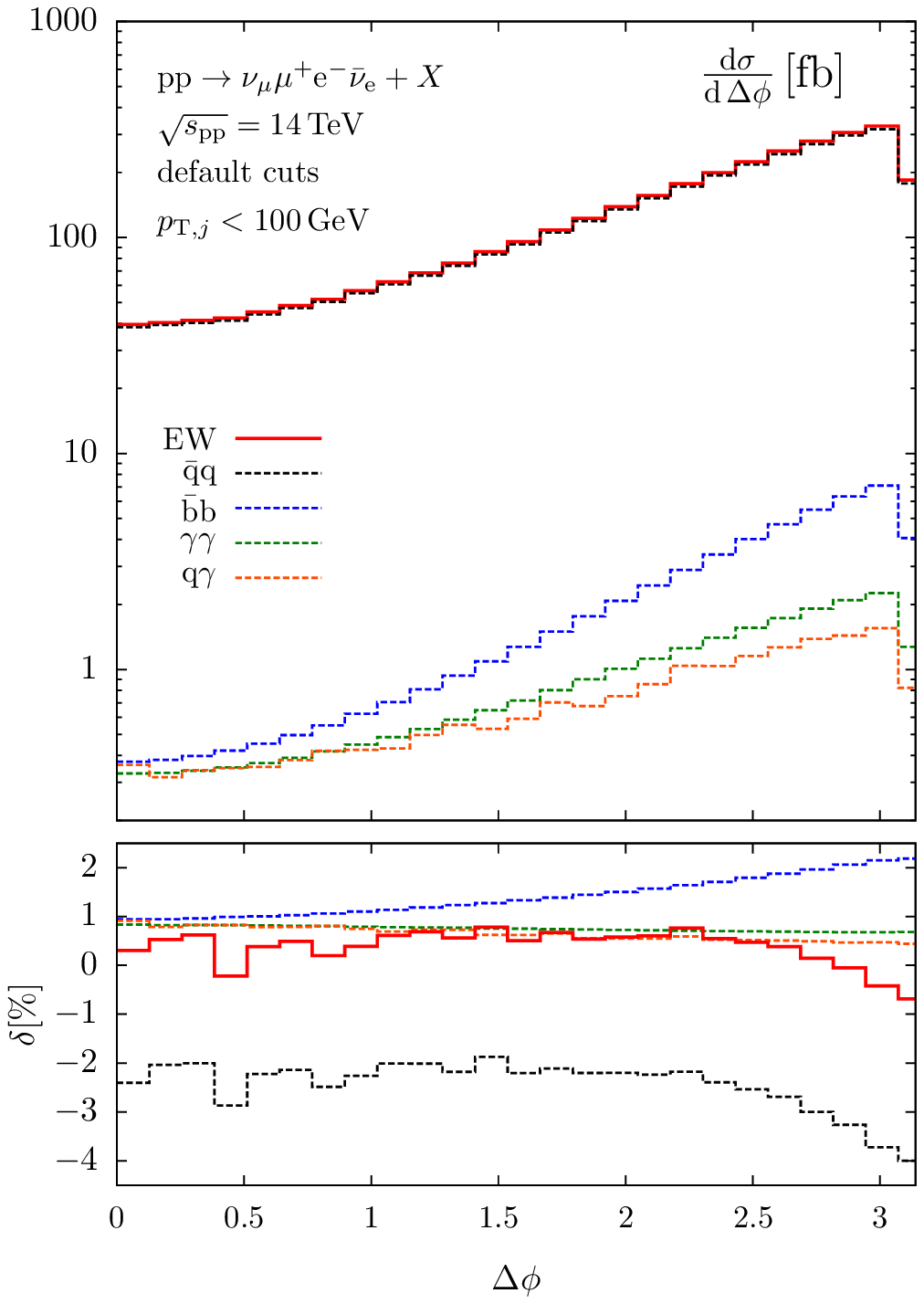}
\caption{Rapidity distribution of the electron (left) and azimuthal-angle separation of the two charged leptons (right) in $\Pp\Pp\to\mvev$ at NLO~EW accuracy (upper panels) for our default setup at the LHC14, together with the relative impact of individual contributions in each case (lower panels). 
\label{fig:angular}
}
\efi
%
%
\bfi
\includegraphics[angle=0,scale=0.7,bb=60 330 370 770]{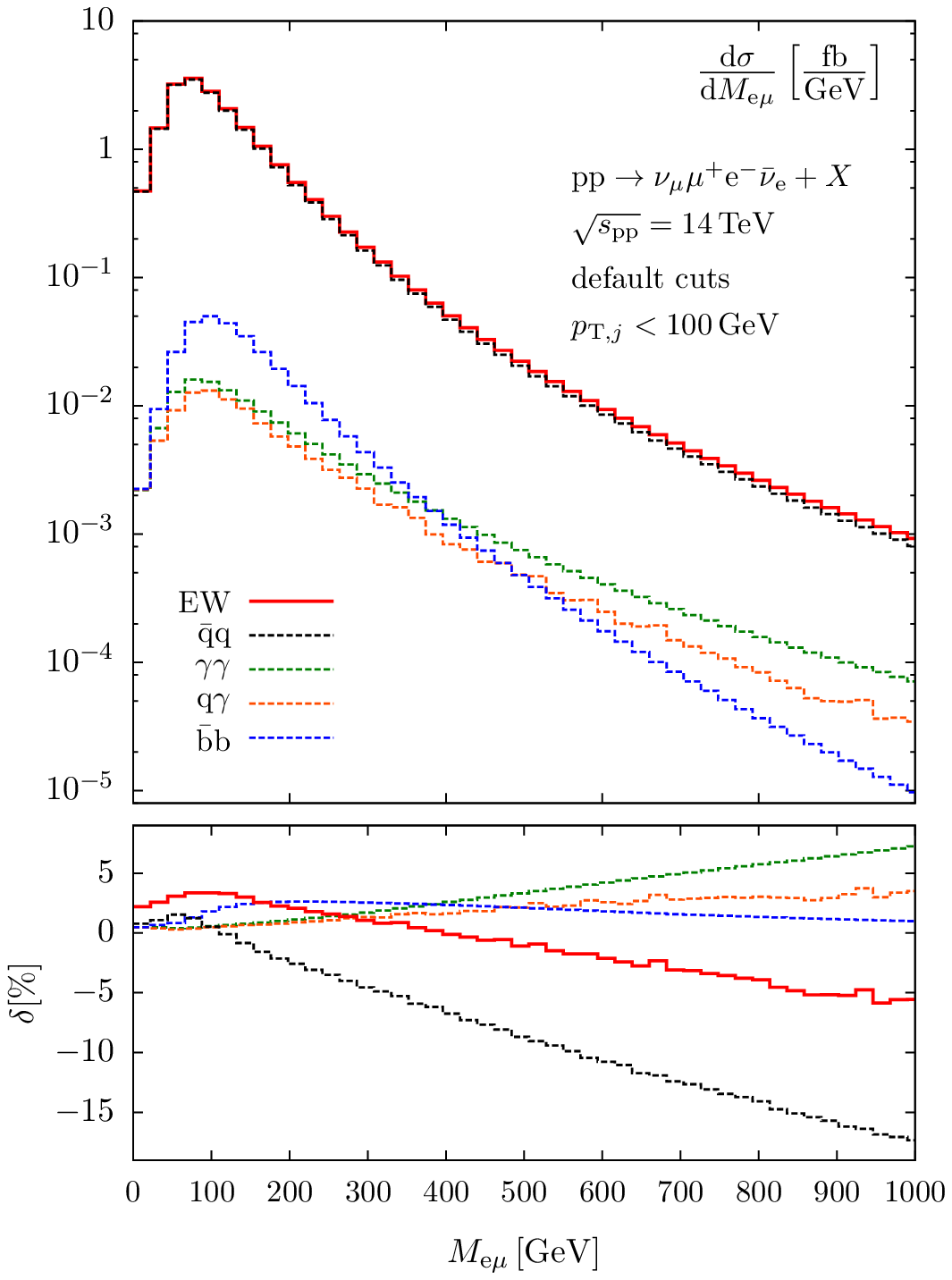}
\includegraphics[angle=0,scale=0.7,bb=60 330 370 770]{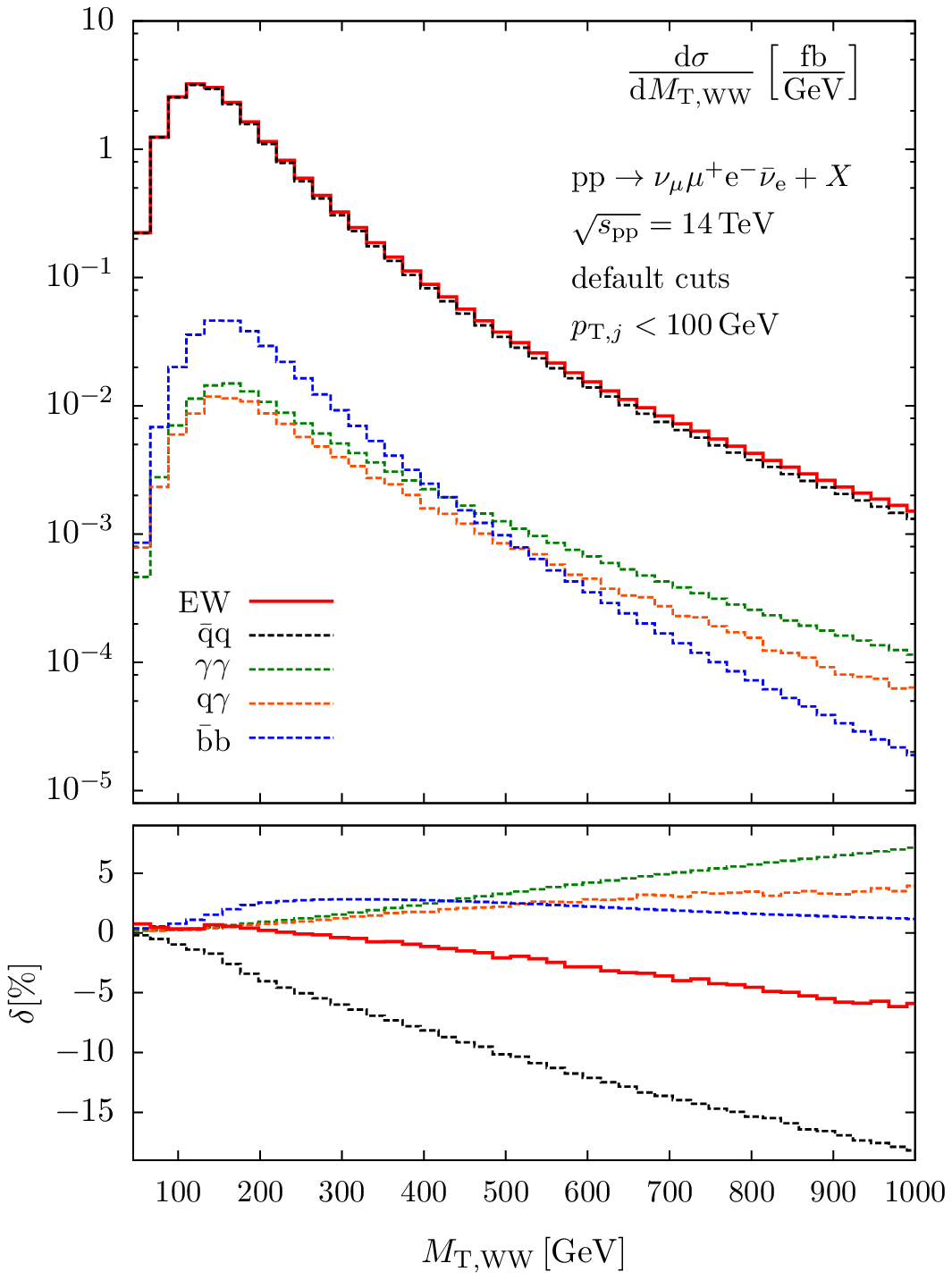}
\caption{Invariant mass of the electron--muon system (left) and transverse-mass distribution of the four-fermion decay system (right) in $\Pp\Pp\to\mvev$ at NLO~EW accuracy (upper panels) for our default setup at the LHC14, together with the relative impact of individual contributions in each case (lower panels). 
\label{fig:mdecay}
}
\efi
%
%
\bfi
\includegraphics[angle=0,scale=0.7,bb=60 330 370 770]{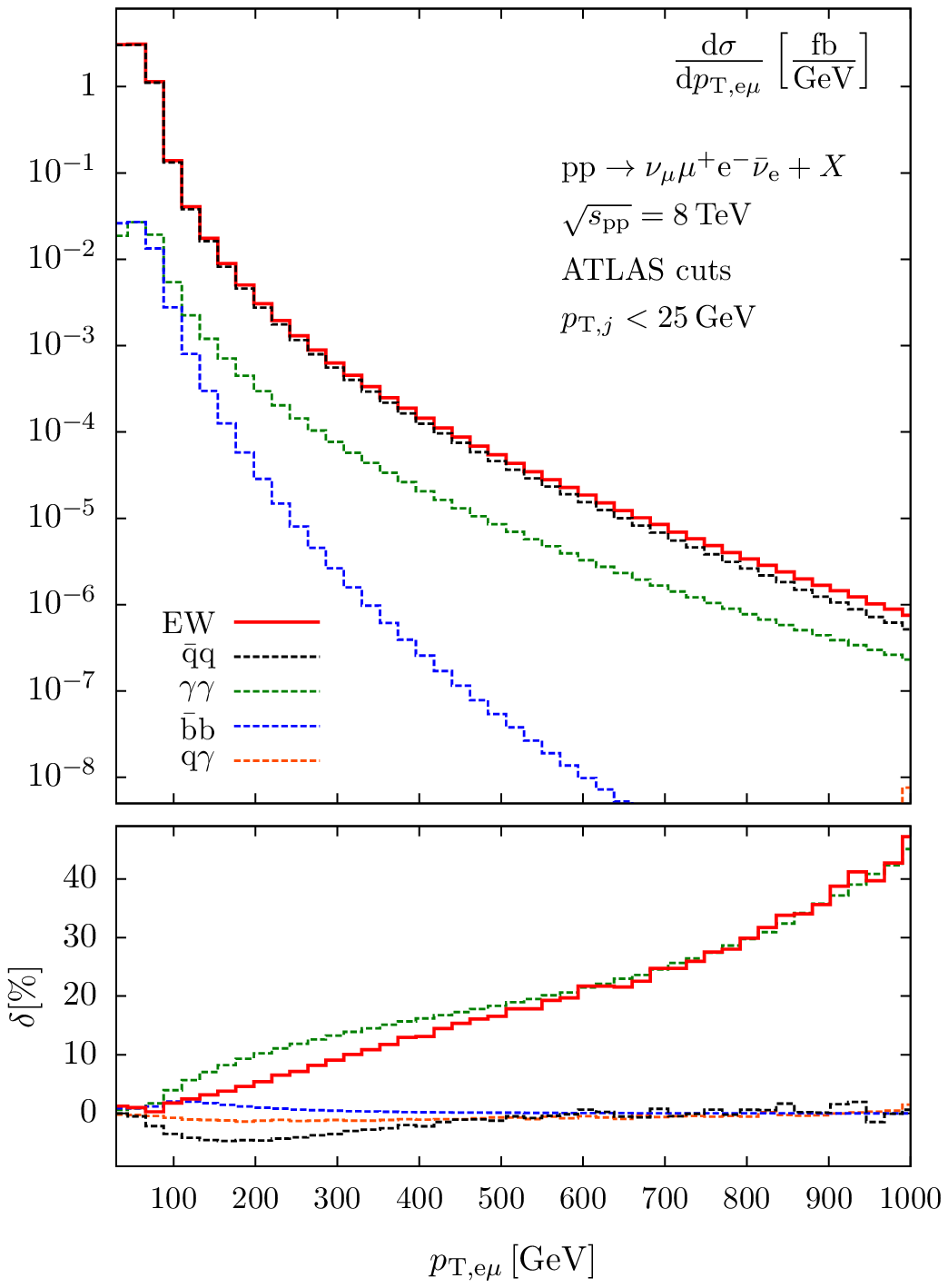}
\includegraphics[angle=0,scale=0.7,bb=60 330 370 770]{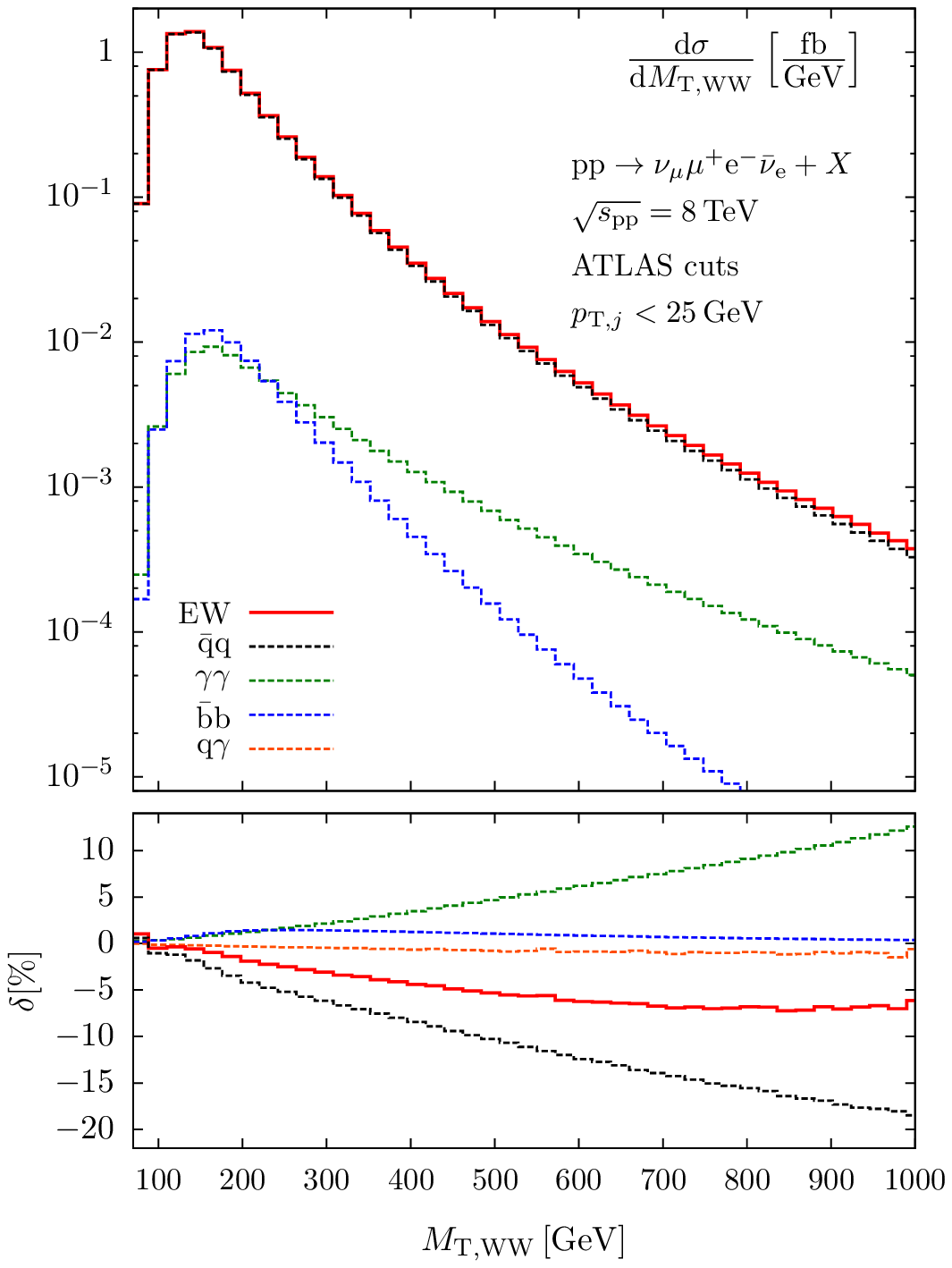}
\caption{Transverse-momentum distribution of the electron--muon system (left) and transverse-mass distribution of the four-fermion decay system (right) in $\Pp\Pp\to\mvev$ at NLO~EW accuracy (upper panels) with realistic selection cuts at the LHC8,  
together with the relative impact of individual contributions in each case (lower panels). 
\label{fig:mt_atlas}
}
\efi
%
%
\bfi
\includegraphics[angle=0,scale=0.7,bb=60 330 370 770]{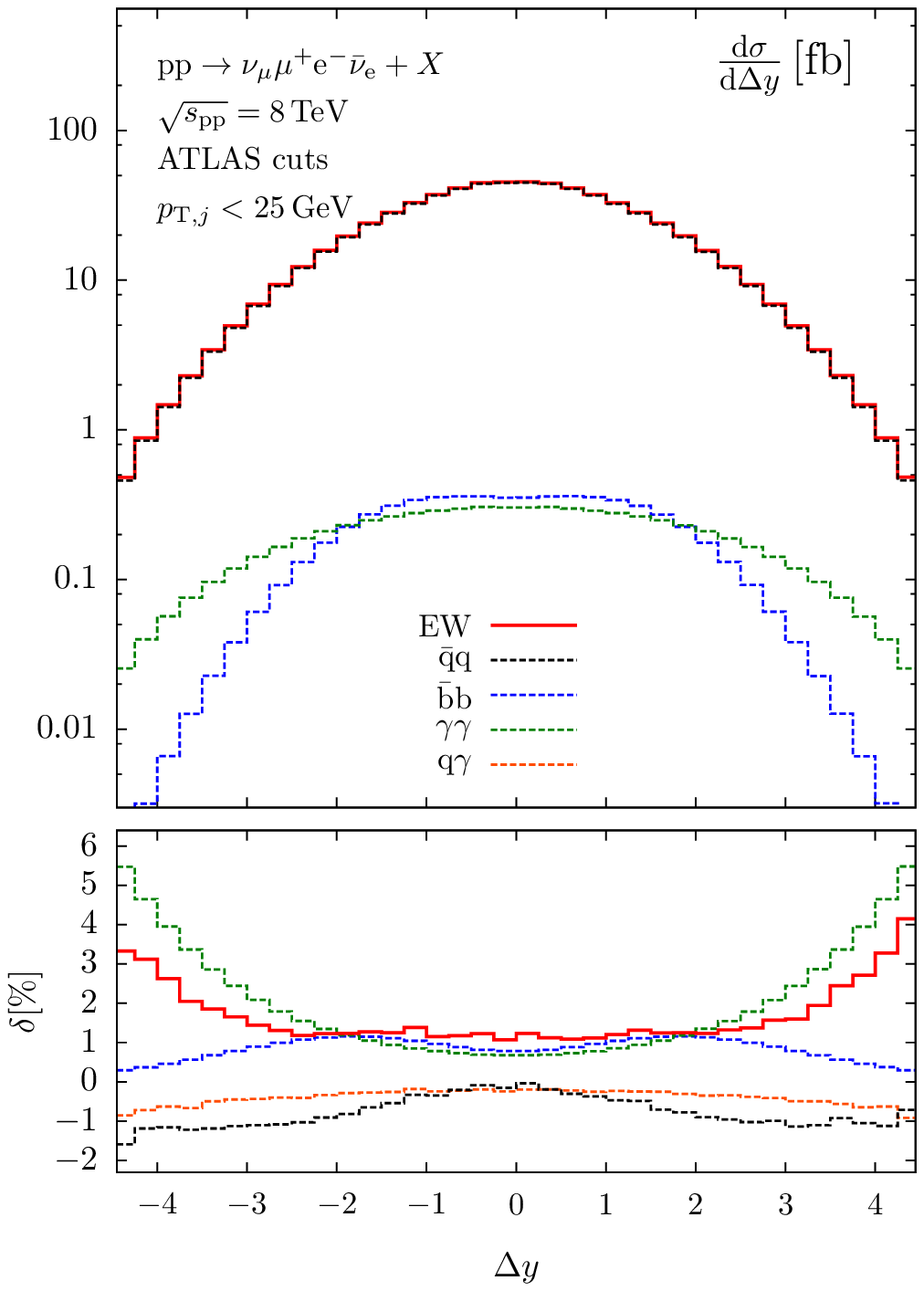}
\includegraphics[angle=0,scale=0.7,bb=60 330 370 770]{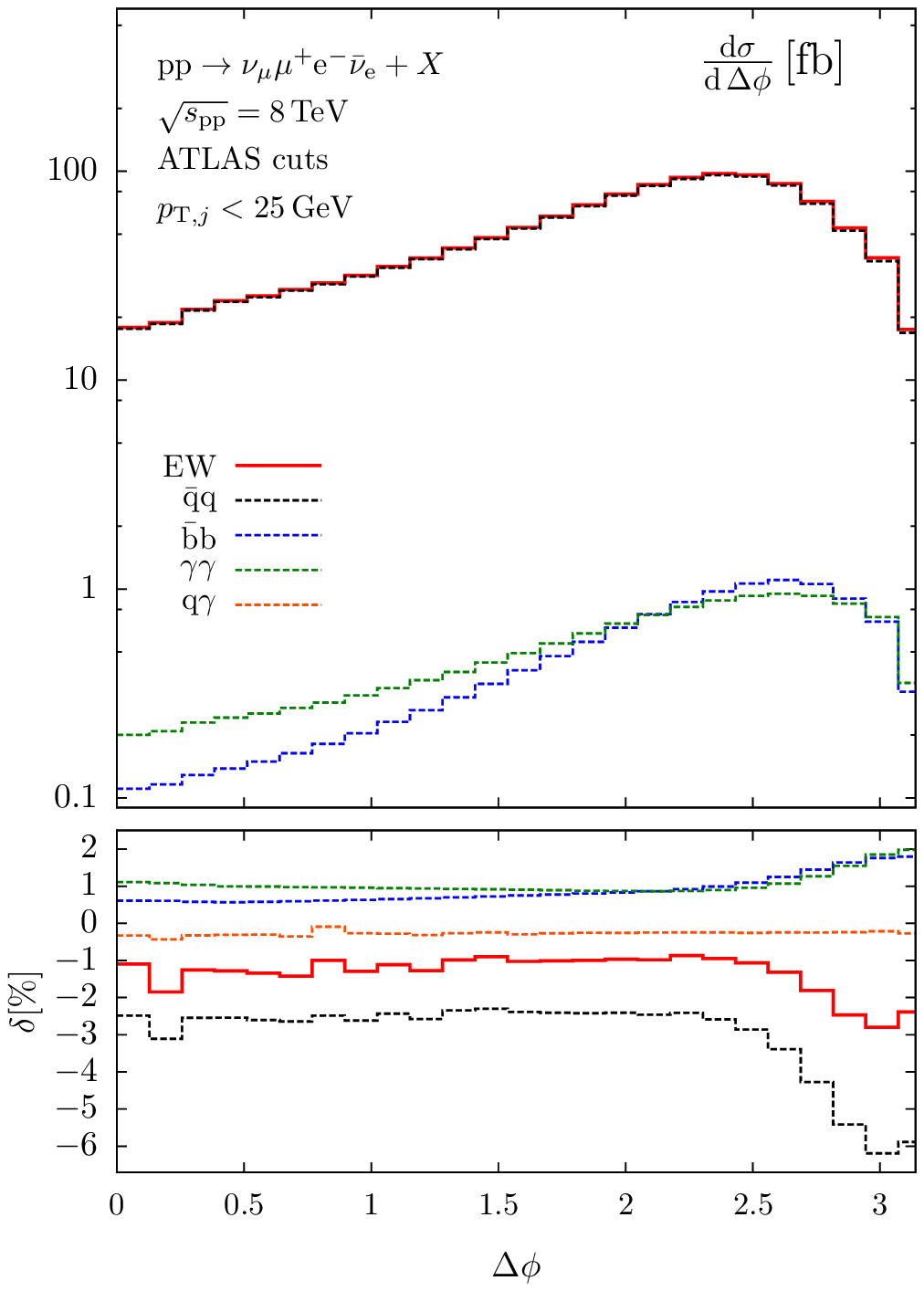}
\caption{Rapidity difference (left) and azimuthal-angle separation of the two charged leptons
(right) in $\Pp\Pp\to\mvev$ at NLO~EW accuracy (upper panels) with realistic selection cuts at the LHC8,  
together with the relative impact of individual contributions in each case (lower panels). 
\label{fig:angular_atlas}
}
\efi
%
In contrast to transverse-momentum distributions, angular distributions of the decay leptons do not experience considerable 
distortions at NLO. 
Radiative corrections mainly change the normalization of rapidity and azimuthal-angle distributions. 
In this context we recall that we consider a setup based on dressed leptons throughout, where large photonic corrections due to
final-state radiation are systematically suppressed. We expect, however, that distributions based on the directions of the
leptons only will not be significantly distorted by collinearly enhanced final-state radiation either, since those effects
do not change the directions of the leptons significantly.
Figure~\ref{fig:angular}~(left) 
shows the various corrections to the rapidity distribution of the electron. The impact of the photon-induced contributions on $\rd\sigma/\rd y_\Pe$ is almost negligible. The pure quark channels yield NLO corrections of about $-3\%$ that are uniformly distributed in $y_\Pe$. 
The NLO~EW corrections to the azimuthal-angle separation of the electron and the muon, $\Delta\phi = |\phi_\Pe-\phi_\mu|$, exhibit a similar behaviour,  
c.f.\ \reffi{fig:angular}~(right). 

The experimental signature of a $\PWp\PWm$ event in the fully leptonic decay mode is that of two oppositely charged leptons 
and missing transverse energy ($E_\rT^\mr{miss}$), which results from the undetected neutrinos. 
In order to distinguish $\PWp\PWm$ candidate events from a priori large background contributions due to Drell--Yan, $\Pt\bar\Pt$, 
and single-top production processes, typically cuts are applied that depend on the values of the invariant mass of the two charged leptons, $M_{\Pe\mu}$, and the missing transverse energy. Qualitatively, the invariant lepton-pair mass, shown in \reffi{fig:mdecay}~(left),
behaves similarly to the transverse-momentum distribution of the electron discussed above. With the jet veto of Eq.~(\ref{eq:jet-veto}), the impact of the quark--photon induced channels on the distribution is marginal. Interestingly, the $\gamma\gamma$-induced tree-level contributions have considerable impact on the tail of $\rd\sigma/\rd M_{\Pe\mu}$, as already observed in
\citeres{Bierweiler:2012kw,Baglio:2013toa} for the related case of on-shell 
W-boson pair production. Our results confirm that those large photonic contributions
survive the application of realistic cuts imposed on the W-boson decay products.
With the largest 
cross-section contribution coming from the $\bar qq$ modes, however, the slight rise of the invariant-mass distribution with increasing values of $M_{\Pe\mu}$ due to the subprocesses with incoming photons is overwhelmed by the large negative corrections caused by the quark-initiated contributions. 

This pattern can also be observed in the transverse-mass distribution of \reffi{fig:mdecay}~(right). 
Since the full invariant mass of the $\PWp\PWm$ system cannot be reconstructed experimentally, it is common to consider the transverse mass, constructed from the transverse momenta of the final-state charged leptons and the missing transverse momentum of each event, 
\beq
M_{\rT,\PW\PW} = \sqrt{
(E_\rT^{\Pe\mu}+E_\rT^\mr{miss})^2
-
(\vec{p}_\rT^{\;\Pe\mu}+\vec{p}_\rT^\mr{\;miss})^2
}\,,
\eeq
with 
\beq
E_\rT^{\Pe\mu} = \sqrt{(\vec{p}_\rT^{\;\Pe\mu})^2+M_{\Pe\mu}^2}, \qquad
E_\rT^\mr{miss} = |\vec{p}_\rT^\mr{\;miss}|. 
\eeq
Here, $\vec{p}_\rT^{\;\Pe\mu}$ denotes the transverse momentum of the $\Pe\mu$~system, and $\vec{p}_\rT^\mr{\;miss}$ the 
total transverse momentum of the neutrino system. 

We consider this distribution in our more realistic setup ``ATLAS cuts'' defined in the previous section.
In combination with the jet-identification criterion of Eq.~(\ref{eq:jet-pt}) the jet veto implies that quark--photon induced events can only contribute to the zero-jet cross section. 
This feature is illustrated by \reffi{fig:mt_atlas} for the transverse momentum of the charged-lepton system and the transverse mass $M_{\rT,\PW\PW}$.
Contributions from the quark--photon channels have been effectively eliminated by the jet veto of Eq.~(\ref{eq:atlas-jveto}).
While in the transverse mass distribution the bulk of the NLO EW 
cross section is now dominated by the pure quark channels, with additional non-negligible  corrections due to the photon--photon induced subprocesses,
the NLO EW corrections to the transverse momentum of the charged lepton system show a completely different behaviour. 
For this observable the negative EW corrections to the $\bar q q$-induced channels are compensated by the large impact 
of the process $\Pp\Pp\to\PWp\PWm\ga$ with real radiation of a hard photon, 
since a large photon recoil allows for
higher values of the transverse momentum of the W~pair which is effectively transferred to the decay leptons
due to the strong boost of the decaying W~bosons.
It turns out that in the current setup 
the photon--photon induced contribution dominates our predictions for the total correction.

Our results on angular distributions are presented in \reffi{fig:angular_atlas}. For the charged lepton rapidity
difference 
$\De y=y_\mu-y_\Pe$ (l.h.s.) as well as for the azimuthal-angle separation (r.h.s.) the corrections slightly increase for
back-to-back configurations of the two charged leptons. While the photon--photon induced contribution dominates the forward--backward 
emission of the charged leptons, the $\bar q q$-induced contribution becomes more important for the back-to-back configuration in the
transverse plane. In both cases the sum over all contributions remains small, leading to total corrections of less than $5\%$.    

\subsection{Scale dependence}

Theoretical uncertainties of a perturbative calculation are often estimated by evaluating cross sections and distributions for different values of the unphysical scales entering the calculation. Since the reaction $\Pp\Pp\to\mvev$ is a pure electroweak process at tree level, the associated LO cross section does not contain any powers of the strong coupling $\alpha_s$, and therefore does not exhibit any dependence on a renormalization scale associated with the running of $\alpha_s$. Clearly, this feature is not altered by the NLO 
EW corrections. However, LO and NLO~EW results do depend on a factorization scale $\muf$ via the parton distribution functions of the scattering protons.  Indeed, we find that varying 
$\muf = \xi \MW$ in the range $0.5<\xi<2$ 
changes inclusive cross sections for our default LHC14 setup 
by about $\pm 8$\% both at LO and NLO in $\alpha$.
The choice of scale does not only affect 
the normalization, but also the shape of kinematic distributions. In \reffi{fig:scale}
%
\bfi
\begin{center}
\includegraphics[angle=0,scale=0.7,bb=60 330 370 770]{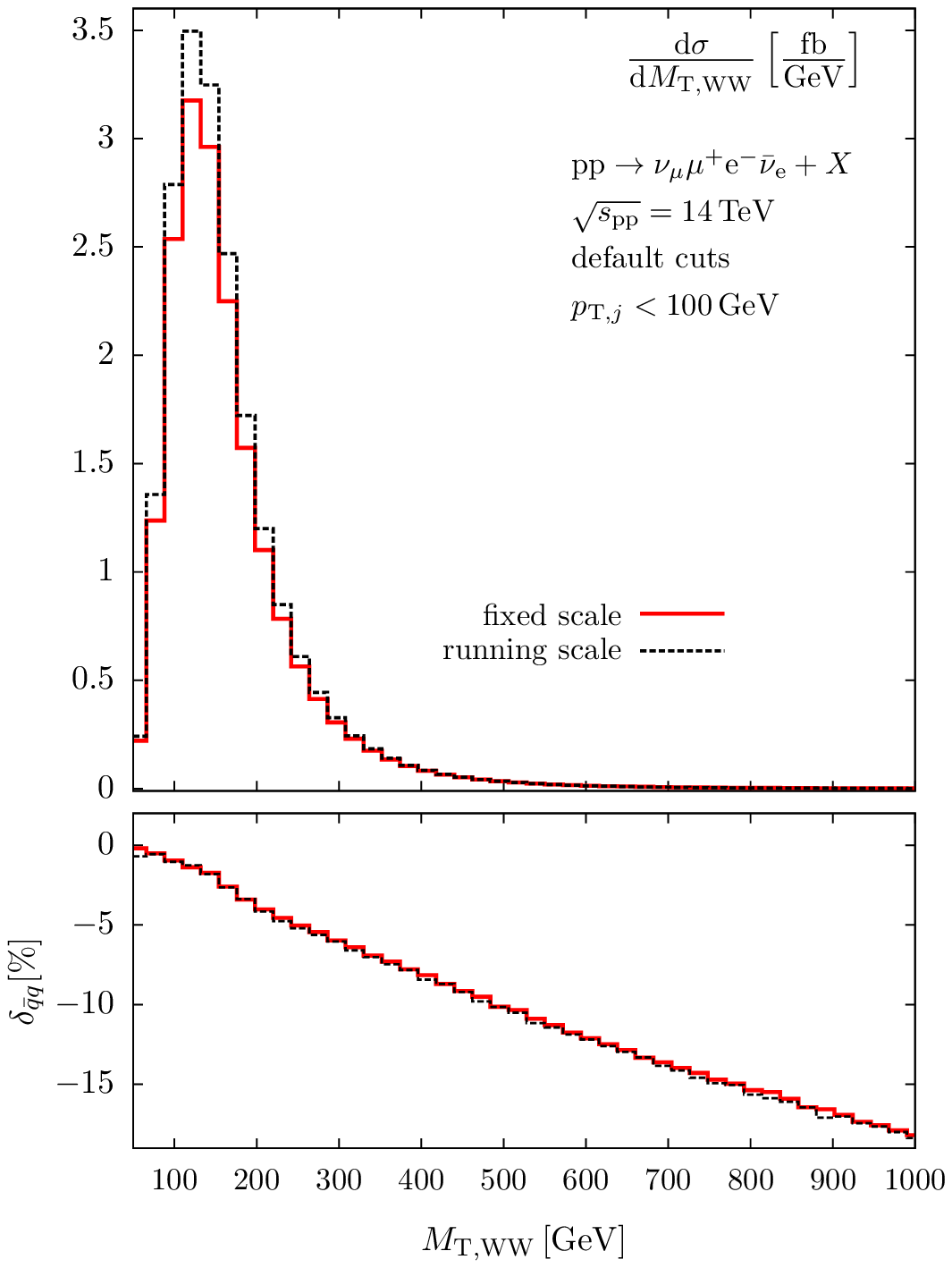}
\end{center}
\caption{Transverse-mass distribution of the four-fermion decay system in $\Pp\Pp\to\mvev$ at NLO~EW accuracy for our default setup at the LHC14 with two different choices of the factorization scale (upper panel), together with their ratio to $\rd\sigma^\mr{LO}_{\bar q q}/\rd M_{\rT,\PW\PW}$  (lower panel). 
\label{fig:scale}
}
\efi
%
this effect is illustrated for $\rd\sigma/\rd M_{\rT,\PW\PW}$, which has been evaluated for our default choice, $\muf=\MW$, and for the factorization scale being identified with the invariant mass of the $\PW\PW$ system, $\muf=M_{\PW\PW}$. For this distribution, 
setting $\muf=\MW$ results in slightly smaller values at low $M_{\rT,\PW\PW}$, but larger values  in the tail. 
Note, however, that the major part of the observed factorization scale dependence is due to the PDFs and will be partly compensated once the NLO QCD correction is added (not considered in this paper). The scale dependence practically disappears in the relative EW corrections, where the PDF effect is mostly cancelled in the ratio $\sigma^{\NLO\,\EW}/\sigma^{\LO}$ if numerator and denominator are evaluated with the same PDF set, as shown in the lower plot of \reffi{fig:scale}.

The small sensitivity 
of the NLO EW corrections to the choice of the PDFs and to the scale settings suggests a simple combination with state-of-the-art QCD predictions 
in terms of the following factorization ansatz,
\beq
\rd\sigma = \rd\sigma^\QCD_{qq}\times(1+\delta^\EW_{qq})+\rd\sigma_{\Pg\Pg}+\rd\sigma_{\ga\ga}+\rd\sigma_{\ga q}.
\eeq
Here $\rd\sigma^\QCD_{qq}$ is the state-of-the-art QCD prediction for W-pair production via quark--antiquark annihilation,
and the remaining additive contributions originate from gluon--gluon, photon--photon, and quark--photon scattering.
This ansatz is particularly motivated by the factorization of the large electroweak logarithmic correction at high energies, which is observed
in the Sudakov regime~\cite{Fadin:1999bq}, but also 
expected in regions such as the Regge limit of forward scattering.
Similar procedures were also adopted in related processes of weak-gauge-boson~\cite{Denner:2011vu} and Higgs production (see e.g. treatment of VBF and $V\PH$ production processes in~\cite{Dittmaier:2011ti,Dittmaier:2012vm}). 

\subsection{Comparison to existing work}

We have compared the results of our calculation to the ones on 
on-shell $\PWp\PWm$ production at the LHC presented in~\citeres{Bierweiler:2012kw,Baglio:2013toa}.
To this end, we consider $\PW$-boson momenta as defined in \refeq{eq:Wmom} with a possible $\ga$ contribution from photon recombination.
Based on these reconstructed momenta we adopt the setup including default cuts of \citere{Bierweiler:2012kw} and
compare to the numerical results presented therein.  
Since we do not further restrict our calculation in DPA the relative EW correction to the leptonic $\PW$ decay
has to be subtracted from our result for a proper comparison. We calculated them to be $-0.66\%$ in total.
Taking this effect into account our relative corrections to the integrated cross sections, $\delta_{\ga\ga}$ and $\delta_{\bar q q}$, agree within $0.6\%$
with the results presented in Table 1 in \citere{Bierweiler:2012kw}. We find even better agreement of $0.2\%$ if we compare totally inclusive cross sections (Table 2 in \citere{Bierweiler:2012kw}). Comparing transverse momentum and rapidity distributions of the W~boson (Fig.~8 and 10 in \citere{Bierweiler:2012kw}) we again find very good agreement within $1\%$ over the 
entire investigated range, while
for the invariant mass of the $\PW\PW$ system (Fig.~9 in \citere{Bierweiler:2012kw}) we observe larger deviations of $3\%$.
Generally we expect larger deviations in observables that are more sensitive to the cuts on the $\PW$-boson momenta,
because our reconstructed $\PW$ momenta can contain contributions from photons resulting from final state radiation, which
are not present in the more idealized calculations based on stable on-shell W~bosons.
Hard cuts on the invariant mass of the $\PW$ pair lead therefore to larger deviations of up to $10\%$ in the $\Delta y$ distribution (Fig.~13 in \citere{Bierweiler:2012kw}).
However, we confirm the observation of large corrections by $\ga\ga$ induced processes in this case.
The results presented in \citere{Baglio:2013toa} can be compared to our relative corrections $\delta_{\ga\ga}$, $\delta_{\ga q}$, and to the combination $\delta_{\EW}-\delta_{\bar\Pb\Pb}$.
After adopting the setup of \citere{Baglio:2013toa} 
we find very good agreement within $1\%$ for $p_{\rT}$ and $y_{\PW\PW}$ distributions (left panel of Fig.~15 and upper right panel of Fig.~16 in \citere{Baglio:2013toa}) and again a slightly larger deviation of $3\%$ when we compare
the invariant mass of the $\PW$ pair (right panel of Fig.~15 in \citere{Baglio:2013toa}).

From this comparison we conclude that the effect of realistic cuts on the $\PW$-boson decay products (and corrections to the decay)
can change EW corrections to the $\PW$-pair production core process by several percent.
%
%
\section{Summary and conclusions}
\label{sec:concl}
In this paper we have 
presented a calculation of the next-to-leading order electroweak corrections to $\PW$-boson pair production at the LHC, taking off-shell effects and spin correlations of the $\PW$ bosons and their leptonic decay products into account in the framework of a double-pole approximation. 
In detail,
our calculation of leading-order and photonic real-emission cross sections is based on full matrix elements with four-fermion final states, while
we have performed a systematic expansion of the virtual electroweak corrections
about the resonance poles of the two $\PW$~bosons allowing us to group them into factorizable corrections to only the production of two on-shell $\PW$~bosons or their decays, and non-factorizable corrections where a soft photon is exchanged between production and decay, or between the two decay subprocesses. 
Photon--photon and photon--quark 
induced subprocesses have been taken into account as well. 

We have 
performed detailed phenomenological studies for a representative setup at the LHC that qualitatively confirm
earlier results obtained for on-shell gauge-boson pair production~\cite{Bierweiler:2012kw,Baglio:2013toa}. Our calculation, however, goes much beyond these works as it allows to impose realistic acceptance cuts on the leptons that are observed in experiment. 
We find that the NLO~EW corrections are dominated by the $\bar q q$-initiated contributions and can reach several tens of percent in some kinematic regimes. In particular, tails of 
transverse-momentum and invariant-mass 
distributions receive large negative corrections because of  electroweak high-energy 
logarithms that grow when the scales in a problem become large compared to the mass of the weak gauge bosons. 
At very high scales, the relative size of the $\gamma\gamma$ channel increases, reaching up to 10\% of the $\bar q q$ tree-level cross section in the TeV range. Since new physics is most likely to manifest itself in that kinematic regime, it is important to keep this purely electroweak effect in mind when interpreting data in the context of searches for physics beyond the Standard Model. 
At first sight, photon--quark induced channels impact some high-energy observables at the same level as the $\gamma\gamma$ channel, but whenever the photon-quark channels grow large,
they get swamped by huge QCD corrections.  
Thus, avoiding unwanted 
QCD corrections by 
a jet veto that removes 
events with a very hard parton in the final state, which is typically related to configurations involving the emission of a soft $\PW$~boson, widely suppresses also
the impact of photon--quark channels.

The calculation described in this paper generalizes recent evaluations of electroweak corrections to W-pair production at the LHC upon including off-shell effects
of the W~bosons and their leptonic decays, allowing for the first time the assessment of electroweak corrections to this process in the presence of realistic event-selection
cuts. 
A comparison to the existing on-shell calculations reveals that realistic cuts on W~decay products rather than on idealized 
stable W~bosons typically change the EW corrections by some percent.
The future inclusion of the presented results in data analysis will certainly reduce 
the uncertainty of predictions due to unknown higher-order electroweak effects
well below the uncertainties inherent in the treatment of QCD effects.

We plan to further refine and generalize the calculation in various directions, such as the treatment of ``bare'' leptons in addition to the currently used
prescription of dressing leptons with collinear photons, the combination of electroweak and QCD corrections,
the inclusion of anomalous triple gauge-boson couplings, and the extension to hadronically decaying
W~bosons. Finally, the structure of the off-shell calculation is deliberately chosen in such a way that we can generalize it from the resonance expansion to the
case of fully off-shell W~bosons with not too much effort, since only the virtual correction require modification.

 %
%
\section*{Acknowledgements}
We are grateful to S.~Forte and U.~Klein for valuable discussions of PDF issues and the concept of dressed leptons in ATLAS analyses, respectively.
The work of M.B.\ and B.J.\  is supported in part by the Cluster of Excellence {\em Precision Physics, Fundamental Interactions and Structure of Matter (PRISMA)},  the Research Center {\em Elementary Forces and Mathematical Foundations} of the Johannes Gutenberg 
University Mainz, by the German Research Foundation (DFG), and the German Federal Ministry for Education and Research (BMBF). 
The work of S.D.\ is supported by the DFG via grant DI 784/2-1. 

%
%

\end{document}